# Impacts of Extreme Heat on Labor Force Dynamics

Andrew Ireland, David Johnston, Rachel Knott

Centre for Health Economics, Monash Business School

February 8, 2024

We use daily longitudinal data and a within-worker identification approach to examine the impacts of heat on labor force dynamics in Australia. High temperatures during 2001–2019 significantly reduced work attendance and hours worked, which were not compensated for in subsequent days and weeks. The largest reductions occurred in cooler regions and recent years, and were not solely concentrated amongst outdoor-based workers. Financial and Insurance Services was the most strongly affected industry, with temperatures above 38°C (100°F) increasing absenteeism by 15 percent. Adverse heat effects during the work commute and during outdoor work hours are shown to be key mechanisms.

___________________________________________________________________

Corresponding author: Andrew.Ireland@monash.edu; Level 5, Building H, Monash University Caulfield, Victoria 3145, Australia

We thank seminar participants at Monash University and conference participants at the 43rd Australian Health Economics Society Conference, the 31st Australian Labour Market Research Workshop, the 2023 Econometric Society Australasian Meeting, 37th PhD Conference in Economics and Business and the 2023 Monash Environmental Economics Workshop for discussions and feedback. This research uses survey microdata collected by the Australian Bureau of Statistics, however the findings and views reported in this publication are those of the authors.

# I. Introduction

Work attendance and hours worked may be sensitive to temperature (Graff Zivin & Neidell, 2014). People may stay home on hot days to avoid working in unpleasant or unsafe conditions (Dillender, 2021). Relatedly, on hot days, people may experience a heat-related illness that requires them to leave work early and be absent in subsequent days (Ireland et al., 2023). Labor demanded by firms may also be temperature-sensitive. Firms may shut down or alter staff requirements on hot days due to equipment breakdowns, occupational health regulations, and changes in the demand for their product or service (Behrer & Park, 2017). Notably, the magnitude of any heat effect is likely context-specific, varying across industries, occupations, locations, and time, depending upon workers' exposure and sensitivity to heat (Lai et al., 2023). There may also be significant temporal dynamics, where work time is reallocated in preparation for or following extreme heat. A complete understanding of these contextual effects and labor force dynamics is essential for policy-makers and firms to develop targeted workplace programs and policies that address the specific challenges faced by different industries and locations.

Evidence on these issues from high-income countries is relatively limited and includes highly variable results, underlining the complex relationships and need for further research. Graff Zivin and Neidell (2014), Dillender (2021), and Neidell et al. (2021) present evidence on the effect of high temperatures on work hours in the United States. Graff Zivin and Neidell (2014) find a negative effect for outdoor workers, but not indoor workers. Dillender (2021) finds no effect for outdoor or indoor workers, though there are significant effects for a subset of usually cold areas. Neidell et al. (2021) find no effect for outdoor workers across 2003–2018, but negative effects for some subsets of years (2003–2007 and 2015–2018).

Complementing this research are studies from low- and middle-income countries, which also provide mixed results. Garg et al. (2020) examine the relationship between temperature and work time in China and find that higher temperatures reduce weekly hours worked for some workers, depending on gender and type of work. Surprisingly, male farmers, who are hypothesized to be at higher risk, are not significantly affected. Somanathan et al. (2021) find that for manufacturing workers in India, hot days (> 35°C) do not contemporaneously increase absenteeism for most workers, but there is some increased absenteeism following hot weeks.[1] Rode et al. (2023) use pooled data for seven countries (Brazil, France, India, Mexico, Spain,

---

[1] In Somanathan et al. (2021), the only significantly positive contemporaneous heat effect (> 35°C) on absenteeism is found for garment workers with paid-leave in climate-controlled plants. No significant contemporaneous heat effects were found for steel and weaving workers, nor for garment workers in non-climate controlled plants.

United Kingdom, and United States) and find effects on work hours similar to those in Graff Zivin and Neidell (2014), with negative effects for "weather-exposed industries" (i.e. agriculture, mining, construction, and manufacturing) and near-zero effects for other industries.[2]

We add to these literatures by using a longitudinal data set of Australian workers spanning 2001–2019 that provides 9.1 million daily responses on work attendance from 649,740 workers. Using within-individual variation in exposure to extreme heat for identification, our main finding is that extreme temperatures (daily maximum ≥38°C or 100°F) cause a 5% increase in work non-attendance (absenteeism), compared to days with temperatures between 26–30°C (79-86°F). Similarly, workers reduce work hours by an estimated 46 minutes in weeks with an average daily maximum temperature >38°C, relative to 26–30°C. Results from an event-study approach show that workers do not make up for this lost time in surrounding weeks, potentially leading to heat-related output losses.

The effects are almost twice as large for outdoor-based workers than indoor-based workers (combined across all industries), but Financial and Insurance Services is the most negatively affected industry. Maximum temperatures ≥38°C (100°F) increase work non-attendance (absenteeism) for finance and insurance workers by 15%, compared to temperatures 26–30°C. Finance and insurance workers usually have paid leave entitlements and high work flexibility, but analysis suggests that long commute times, often via public transport, is more important for explaining this result. These heterogeneity results highlight that dividing the sample *a priori* between "high-risk" and "low-risk" industries can conceal important effects.

We also focus on adaptation by exploring variation in effects across regions and time. Large effects are found in areas with historically fewer extreme heat days, suggesting long-run adaptation. However, large effects are also found for the most recent time period (2013–2019), which had the highest number of extreme heat days, suggesting difficulties in adapting to increased heat in the short run.

The remainder of the paper is organized as follows. Section II and Section III explain the data and empirical strategy. Section IV reports our main results on work attendance and hours worked. We then investigate the dynamic and temporal effects of temperature in Section V to determine if lost time is offset by increases in other weeks. We further extend the analysis by exploring heterogeneity in effects in Section VI and adaptation in Section VII. Finally, we conclude with a discussion on the implication of our results in Section VIII.

---

[2] A related literature explores the impacts of heat on productivity. Sahu et al. (2013) show that high temperatures induce cardiac strain in agricultural workers, which reduces productivity. Deryugina & Hsiang (2014) find that heat has a negative and roughly linear impact on productivity. Factory output in India (Somanathan et al., 2021) and China (Zhang et al., 2018) have been shown to reduce with extreme heat, with effects substantially mitigated by climate control. Heyes & Saberian (2022) find that in a sample of Indian workers, each day above 37.7°C (100°F) in a month increases self-reported "inability to work" by 7% for the month.

## II. Data

### A. The Longitudinal Labour Force Survey

Data come from the nationally representative Longitudinal Labour Force Survey (LLFS), which includes individual-level demographic and work-related information from 26,000 randomly selected households across Australia each month. All individuals in the household aged over 15 years are surveyed, covering 0.32% of the civilian population.[3] The survey follows selected households over eight consecutive months, with one-eighth of the sample replaced monthly.

In each survey, respondents are asked about their work activity during a reference week, including the specific days they worked, the number of hours they worked, and reasons for working fewer hours than usual. Information on whether respondents worked on each of the seven days in the reference week allows us to match work attendance to weather conditions on a specific day. Information on work hours is only recorded for the entirety of the reference week, and so we match work hours to weather conditions for the week.

The LLFS uses Statistical Area Level 4 (SA4) structure of the Australian Statistical Geography Standard as its main geographic variable.[4] SA4s are designed specifically for the labor force survey to reflect labor markets within states. We restrict the sample to the 45 SA4s (from the total 88 SA4s in Australia) that experience extreme temperatures and allow more precise temperature measurements (see Appendix A). These 45 areas account for 67% of the Australian labor force. The median number of workers in each included SA4 is 171,317, based on official residential population estimates for 2019, and the median area is 1,032km$^2$ (equivalent to a circle with radius 18.1km or 11.2 miles). Further details of the SA4 inclusion criteria are provided in Appendix A.

---

[3] Surveys are typically conducted in the two weeks beginning on the Sunday between the 5th and 11th each month (see Appendix B, Figure B1). To account for the large number of planned absences and work place closures over the Christmas and New Year period, survey dates for December and January are modified (Australian Bureau of Statistics, 2022). Interviews start between the 3rd and 9th December and the 7th and 13th January. The number of responding households varies between 20,836 (July 2008) and 33,252 (March 2007).

[4] The SA4 recorded for each individual represents the location of usual residence, which may differ from their place of work. However, using data from the 2016 Census of Population and Housing, we find that the average of the included areas' median commuting distance to work is only 11km, while the median SA4s in our sample spans 1,032km². This implies that most included workers experience a similar temperature at home and at work, given that temperature is highly correlated across space, even up to distances of 1200km (Hansen & Lebedeff, 1987).

Our analysis sample includes data from Australia's five hottest months (November–March) from November 2001 to February 2020, excluding public holidays.[5] We concentrate on hotter months because our aim is to estimate the effects of hot weather on labor force dynamics (though we also provide estimates for cooler months). We further restrict our sample to single job holders who worked at least one day during the reference week. The former restriction is imposed because our data only captures daily work attendance for an individual's main job. The set of days included for each worker are restricted to those days they ever report attending work, across any month (i.e. responses on Sundays are only included for individuals who ever attended work on a Sunday). The combined restrictions mean that in our estimation sample individuals have an average of 13.9 reference days (SD=8.4).

Our final sample comprises 9.1 million responses of work attendance from 649,740 individuals (Table 1). Work attendance, our primary outcome, equals 81.2% on average for all days. Attendance is highest on Wednesdays (88.3%) and lowest on Sundays (44.1%) (see Appendix Figure B2). We observe 1.68 million responses for the week-level variables, such as weekly hours worked. On average, individuals in our sample work 35.2 hours each week. Average attendance and weekly work hours by gender, age and occupation groups are presented in Appendix Table B1.

### B. Climatic Data

Daily weather observations from all Australian weather stations were obtained from the Bureau of Meteorology. The station data contains daily maximum temperature, daily minimum temperature, 3-hourly synoptic temperature and humidity observations, and daily precipitation readings. Throughout the paper, we use daily maximum temperature as our primary measure for heat. We estimate the weather at the location of the population-weighted centroid for each SA4 using inverse distance weighting.[6]

---

[5] We exclude data from March 2020 onwards from our primary analysis due to possibility of confounding factors arising from the onset of the COVID-19 pandemic in Australia and subsequent lock-down policies.
[6] To determine the population weighted centroids for each SA4 we use the 2019 residential population of the smaller Statistical Area Level 2 (SA2). We take the weighted average of the five closest weather stations to the SA4 centroid. Weights are calculated as the by the inverse of the distance squared, giving closer weather stations more weight.

There is considerable variation in climate across Australia (Appendix C), with extreme temperatures most common in the south. The major cities of Perth and Adelaide experience the most days above 38°C: a threshold deemed dangerous in the likelihood of heat disorders, at all humidity levels (NOAA National Weather Service, 2023) (see Appendix Table C1 and Figure C1). We exclude northern regions with tropical and subtropical climates and Tasmania, Australia's southernmost state, which rarely experience extreme temperature highs exceeding 38°C (100°F) (Figure C2). Overall, 3,840 (3.0%) SA4-days in our analysis period, the level of our identifying variation, have a maximum temperature over 38°C (Figure C3). Extreme heat occurs most frequently in January, with 5.9% of the day level responses on days exceeding 38°C (Figure C4).

### C. Employment Characteristics

We use data from O*NET OnLine (National Center for O*NET Development, 2021) and the Household, Income and Labour Dynamics in Australia (HILDA) Survey to enrich our analysis on the likely working conditions and possibilities for heat avoidance and adaptation amongst various occupations and industry subgroups. O*NET contains detailed ratings by occupation for many characteristics. In particular, we use scores for time spent outdoors exposed to the weather to classify outdoor- and indoor-based occupations.[7] This classification leads to 9.2% of responses grouped as "outdoor", since it only includes occupations with the highest likelihood of frequent unprotected outdoor exposure. These include occupations such as beef cattle farmer, glazier, police officer, tennis coach, recycling worker and bulldoze operator.

The HILDA survey has followed a nationally representative sample of approximately 17,000 Australian households since 2001 (Wooden & Watson, 2007). It collects a broad range of information on income, employment, education, health, wellbeing and relationships. For occupation and industry subgroups analyzed in Section VII, we use HILDA responses to measure job characteristics such as work flexibility, wages and the ability to work from home.

---

[7] We find similar effects with a broader industry-based definition of outdoor work aligning with Graff Zivin & Neidell (2014). Outdoor industries by this definition are Agriculture, Forestry & Fishing; Mining; Manufacturing; Electricity, Gas, Water & Waste Services; Construction and Transport, Postal and Warehousing. All other industries are classified as indoor.

# III. Empirical Approach

Our main analysis is conducted at the individual-day level using a binary outcome variable, $Y_{ita}$, denoting work attendance for individual *i*, in area *a*, on day *t*:

$$Y_{ita} = \sum_{j=1}^{J} \beta_j * temp_{j,at} + \phi_{ia} + \tau_t + X'_{ta}\gamma + W'_{ta}\varphi + \epsilon_{ita} \quad (1)$$

Temperature is represented in equation (1) by six binary variables ($temp_{j,at}$) indicating that the daily maximum temperature in area *a* on day *t* was in that category. The categories are defined as <22°C, 22-26°C, 26-30°C, 30-34°C, 34-38°C, and >38°C. The 26–30°C indicator variable is the reference category in each regression.

Individual-area fixed effects ($\phi_{ia}$) are included to account for time-invariant characteristics that vary across individuals and areas, such as differences in usual work attendance.[8] We use individual-area fixed effects, rather than separate individual and area fixed effects, to account for differences in work schedules and temperatures when individuals move locations. We include year-month fixed-effects ($\tau_t$) to account for time-varying factors affecting work attendance common across individuals, such as levels of unemployment. We also include two additional temporal fixed effects, $X'_{ta}$, in our specification: (i) calendar-date fixed-effects to account for date-specific factors that influence attendance, such as days following public holidays or other annual events; and (ii) day-of-week fixed-effects since work attendance is lower on weekend days. Finally, other weather-related variables correlated with temperature are represented by $W'_{ta}$. Specifically, we control for daily precipitation using four categories (≤ 2mm, 2–5mm, 5–10mm and > 10mm) and daily maximum wind gust. Standard errors are clustered at the area level.[9]

Individuals have up to 35 daily attendance observations over the five included months, and there is sufficient within-individual temperature variation over these days to precisely identify the temperature effects ($\beta_j$). Appendix Figure C5 illustrates this variation with a histogram of the within-individual interquartile range. The within-individual-area interquartile range in experienced temperatures across days is greater than five degrees Celsius for 70% of individual-areas and greater than ten degrees for 12%. The temperature variation is centered around an average of 26.7 °C (see Appendix Figure C3)

---

[8] Areas are SA4 geographical units. See Section 2 for further details.
[9] Starndard errors are comparable (slightly smaller) when clustered at the individual and household levels (Appendix D)

A weekly analogue to equation (1) is used to estimate the effects of average daily maximum temperature on the number of hours worked during the week.[10] We use the same 4°C bins as in the day-level analysis, again with 26–30°C as the reference category. Individual-area fixed effects ($\phi_{ia}$) and year-month fixed-effects are included for the same reasons as above. Rather than calendar-day fixed-effects, week-of-year fixed effects account for seasonality. To control for precipitation, we use bins for total rainfall during the week (≤10mm; 10 to 100mm; >100mm). The incidence of public holidays during the week, which are associated with low work attendance, is controlled for with a covariate representing the number of public holidays (public holiday days are excluded from the daily panel).

In addition to these main empirical approaches, several alternative regression specifications are used to test the sensitivity of the estimated effects. Examples include using wet-bulb temperature, controlling for pollution, and including different fixed-effects. We also test for sensitivity by using different estimation samples, such as omitting days with any rainfall, including more and fewer months of the year, restricting to full-time workers, and altering the geographical areas included.

# IV. Main Effects of Temperature on Labor

## A. Temperature and Daily Work Attendance

Estimates of Equation (1) indicate that high temperatures reduce work attendance (Figure 1). Temperatures above 38°C cause a 0.96 percentage point reduction in the likelihood of working (p<0.01), relative to a 26–30°C day (5.1% increase relative to a non-attendance rate of 18.8%). The effect is larger for workers in outdoor occupations (defined using O*NET): temperatures above 38°C cause a 1.87 percentage point reduction in the likelihood of working for outdoor workers (p<0.01) and a 0.82 percentage point reduction (p<0.01) for indoor workers. These effect sizes equate to a 10.0% and 4.3% increase in work non-attendance, relative to sample means, for outdoor and indoor workers, respectively. Notably, hot temperatures of 34–38°C do not affect work attendance (-0.05 percentage points), even for outdoor workers (-0.25 percentage points), suggesting that maximum temperatures must be in the 'extreme' range to have an effect.

---

[10] Our analyses are based on the average maximum temperature calculated across all seven days in the week. Since we can determine each worker's usual work days based, based on their attendance across all survey months, an alternative measure of heat is the average maximum temperature on usual work days. The main effects are similar using this measure (Appendix D, see Figure D1).

Outdoor workers may be more sensitive to heat because many adaptation strategies are unavailable, such as air conditioning (Barreca et al., 2016) and working from home (Li, 2020). Also, some outdoor-based workplaces, such as unionized construction worksites, have heat policies to cease work when temperatures exceed a threshold (CFMEU Victoria, 2015). More generally, the reduced work attendance on days greater than 38°C is unlikely to be explained by increased individual demand for outdoor leisure. Graff Zivin and Neidell (2014) find that extreme heat over 100°F (38°C) reduces work and outdoor leisure activities but increases time spent on indoor leisure activities.

For outdoor workers, we additionally find a reduction in work attendance at low temperatures, though considerably smaller than for temperatures above 38°C. Temperatures below 22°C cause a 0.38 percentage point reduction in the likelihood of working for outdoor workers ($p$=0.02), relative to a 26–30°C day.[11] We primarily focus on the high-temperature effect because in Australia (and most other places in the world), the change in the number of hot days is predicted to be considerably greater than the change in the number of cold days, making it the more important policy issue.

B. Temperature and Work Hours

Figure 2 shows the effect of the average daily maximum temperature during the reference week on the total hours worked in all jobs. We find that in a week with an average maximum temperature over 38°C, workers reduce their work hours by 0.77 hours ($p$<0.01), compared to a week averaging 26–30°C. This equates to a 2.2% reduction in working hours, relative to a sample mean of 35.2 hours per week. Outdoor workers are more sensitive to temperature and reduce their hours by 2.02 in such weeks (5.0%, $p$<0.01). These significant negative effects suggest that workers do not make up for higher absenteeism by working longer hours on cooler days in the same week, though we more-thoroughly test for such dynamics in Section V.

Table 2 shows the effects of high temperatures on total work hours and three additional weekly outcomes: days worked per week, hours worked per day, and an indicator variable denoting a self-reported assessment that the individual worked less than usual. The reduced weekly hours caused by extreme temperatures, illustrated in Figure 2 and Column (1) of Table 2, can result from reduced work days (extensive margin) and reduced work hours when at work (intensive margin). Column (2) and Column (3) of Table 2 explore how much of the reduced weekly hours is caused by these different responses. We find that days per week significantly decrease in weeks of extreme heat, with an average maximum temperature over 38°C causing workers to work 0.06 ($p<0.01$) fewer days per week on average, relative to weeks with average temperature between 26 and 30°C. Average hours worked on working days also decreases when average maximum temperatures exceed 38°C (0.05 $p=0.03$). We calculate that work attendance accounts for 69% of the overall reduction in hours, while the remaining 31% occurs due to shorter work days (for those attending work).[12] This split is similar for outdoor (63%) and indoor workers (71%).

---

[11] Although 22°C is not especially cold, in Australia due to a number of factors, people are more affected by cool weather than in other colder climates (Johnston et al., 2021). The mean maximum temperature on days less than 22°C, for our sample of November–March months, is 19.7°C.

The results we have presented represent the average response across workers. Unsurprisingly, this average effect is driven by a change in work hours among a relatively small group of workers. Estimates in Column (4) show that when the average maximum temperature is over 38°C, the percent of people reporting they worked 'less than usual hours' increases by only 6.47 percentage points (relative to a sample mean of 21.64%). Again, the effect is larger for outdoor workers (14.14) than indoor workers (5.88). The percent of people reporting they worked less than usual is also significantly increased when average maximum temperature is 34–38°C for both outdoor workers (4.33 percentage points $p<0.01$) and indoor workers (1.72 percentage points $p<0.01$).

Survey respondents are asked about their main reason for working less than usual or less than full-time hours.[13] We use the responses to investigate the mechanisms driving the extreme heat effects. Specifically, we estimate six regressions – based on the specification used for weekly work hours – using the following responses as binary outcome variables: (i) own illness, (ii) holiday or personal reasons, (iii) bad weather or plant breakdown, (iv) insufficient work available or stood down (underemployment), (v) shift work or variable hours, and (vi) other reason. We restrict the sample to outdoor workers because the effect of extreme heat on hours worked has an especially large impact for this group. Results for indoor workers are provided in Appendix Figure D2.

---

[12] To decompose the effect, we take the estimate for reduced days per week (column 2) and multiply by the average hours per day (7.75) to estimate the implied average weekly hours lost from reduced work attendance. Similarly, we multiply the estimated decrease in hours per day (column 3) with the average days worked per week (4.50) for the implied weekly hours lost per week from shorter work days. The estimate for hours lost due to work attendnace accounts for 69% of the sum of these two estimates.
[13] Full-time workers who worked less than 35 hours are asked the reason for working part-time hours. From July 2014 onwards, all workers are asked the eason for working fewer hours than usual, when applicable.

Figure 3 illustrates that high temperatures (average weekly temperature above 34°C) cause a significant 1.70 percentage point increase in the likelihood that the respondent worked less because of "bad weather or a plant breakdown". This equates to a 173% increase compared to the mean (0.98). Unfortunately, we cannot disaggregate this outcome, and so we're unable to determine whether the estimated effect is driven by worker behaviour (e.g. avoiding work because of "bad weather") or employer behaviour (e.g. workers sent home because of a "plant breakdown").

The results also indicate that people take significantly more leave during hot weeks (1.08 percentage point increase). This effect may be driven by people avoiding unpleasant work conditions, even though the worksite is still operating normally. The near-zero effect for "illness" is also notable. Research demonstrates that worker health is negatively affected by high maximum temperatures (Dillender, 2021; Ireland et al., 2023), but this adverse effect may be too infrequent to cause measurable effects on weekly work hours.

### C. Robustness of Results

We run several checks to demonstrate the robustness of our results (see Appendix D). First, we find a similar pattern of results if we use wet-bulb temperature, which accounts for humidity (see Appendix, Table D1). This finding is expected given that most geographical areas in our main sample have temperate climates with low humidity during summer.

Second, we test the sensitivity of results to replacing individual-area fixed-effects with area fixed-effects (see Appendix Table D2 column 2). This specification aligns more closely with the approach Graff Zivin & Neidell (2014) used in their analysis of pooled cross-sectional time use data. Estimates are very similar to those from our main specification: temperatures above 38°C are estimated to cause a 0.96 percentage point reduction in the likelihood of working, relative to a 26–30°C day.

In our third robustness model, we include air pollution (particulate matter 2.5) to account for the potential for labor supply to be affected by smoke from wildfires, which are more likely on extreme heat days. Pollution is measured using NASA's MERRA-2 gridded data (Global Modeling Assimilation Office, 2015). Temperature estimates from this model (column 3) are similar to our main estimates. Temperature estimates are also relatively unchanged when restricting the sample to days without rainfall (column 4), implying that our results are not driven by precipitation.

We additionally explore the sensitivity of the estimates to using alternative estimation samples (Figure D3). Using a larger sample of all twelve months or a smaller sample of only Summer months (December, January and February) has little impact on the estimated effect of temperatures above 38°C on work attendance. This insensitivity is due to extreme heat days being concentrated in the summer months, so including additional months doesn't significantly alter our identifying variation. The estimates remain unaffected by excluding responses for each year's first and last week, corresponding to Australia's major holiday season when many workers are absent and workplaces closed. Restricting the sample to full-time workers and weekdays does not alter our main results. Our main estimates are also insensitive to excluding large geographical areas, where the temperature measurement is less precise, and the inclusion of large remote areas, which are omitted from our main sample.

# V. Dynamic and Temporal Effects of Temperature

## A. Prior Heat Exposure

Heat may have lasting effects that impact work attendance on subsequent days. Prior-day extreme heat may increase fatigue and cause illness, reducing today's work attendance, particularly for outdoor workers (Ireland et al., 2023). High nighttime temperatures can also increase fatigue by interrupting sleep (Carias et al., 2022). In Table 3, we investigate these potential lagged effects by comparing the estimated effect of today's maximum temperature (column 1) with the estimated effects of yesterday's maximum daytime temperature (column 2) and last night's midnight temperature (column 3). All three measures are included in the same regression in Column (4). The results indicate that today's temperature (≥38°C) has the largest effect on work attendance decisions, with yesterday's temperature also having a negative effect, but to a considerably smaller extent. For instance, in Column (4) for the all-worker sample, extreme heat today reduces (contemporaneous) work attendance by 0.88 percentage points, while extreme heat yesterday reduces work attendance by only 0.27 percentage points.[14] The effect of midnight temperatures over 26°C is small for outdoor (-0.22, $p$=0.39) and indoor (-0.07, $p$=0.65) workers. These results suggest that lasting heat-induced illness and fatigue (from working in extreme heat yesterday) may play a minor role in work attendance decisions, but that poor sleep is not a key mechanism.

---

[14] While these results may suggest a weak effect for yesterday's temperature, this is not a result of a cumulative heat-effect. We also estimated a regression with an interaction term between today and yesterday and found that the effects are not multiplicative for two consecutive extreme heat days (Appendix E).

### B. Do workers make up for lost time?

In the previous sections, we established that extreme temperatures cause a reduction in both daily work attendance and weekly working hours. The reduction in weekly hours worked implies that workers do not fully make up for lost days within the week. However, workers may compensate for lost working hours in subsequent weeks or even prepare for forecasted hot conditions by altering hours in prior weeks. If workers can adapt to heat by making up for reduced hours or by substituting their days off to work on milder days, the reduced working hours in extreme heat could positively impact firm productivity and workers' safety.

We estimate an event-study regression with indicator variables for average daily maximum temperature ≥38°C for the current week (denoted as week 0), the prior three weeks (-1, -2 and -3), and the subsequent three weeks (+1, +2 and +3) as shown in Equation 2. Individual-area fixed-effects ($\phi_{ia}$), year-month fixed-effects ($\tau_t$), calendar-date and day-of-week fixed-effects ($X'_{ta}$), and weather controls ($W'_{ta}$) are the same as the main analysis.

$$HoursWorked_{ita} = \sum_{k=-3}^{3} \beta_k * Over38°C_{k,at} + \phi_{ia} + \tau_t + X'_{ta}\gamma + W'_{ta}\varphi + \epsilon_{ita} \quad (2)$$

Figure 4 shows the estimated lag and lead temperature effects ($\beta_k$). As shown in previous sections, there is a significant reduction in working hours in the week of high temperatures for both indoor and outdoor workers. Conditional on temperatures in the prior three weeks and subsequent three weeks, working hours reduce by 1.88 ($p<0.01$) and 0.56 ($p<0.01$) for outdoor and indoor workers, respectively, in weeks with average maximum temperature over 38°C, relative to weeks below 38°C. However, the estimated effects of a high-temperature week on work hours in subsequent weeks (corresponding to weeks 1, 2 and 3) are all close to zero and statistically insignificant. This implies that workers do not make up for lost work time. The estimated effects of a high-temperature week on work hours in past weeks (corresponding to weeks -1, -2 and -3) are also close to zero. This implies that workers do not work longer hours in preparation for time off during the forecasted high-heat week.

# VI. Which workers are most affected?

Figure 5 shows the effect of temperature on work attendance, estimated separately by industry.[15] We focus here on work attendance because previous results show that high temperatures reduce attendance more strongly than hours worked per day. Unexpectedly, the results indicate that the largest negative effect occurs for workers in the Financial and Insurance Services industry: a day above 38°C causes a 1.73 ($p$<0.01) percentage point reduction in the likelihood of working, relative to a 26–30°C day. The large effect for this mostly indoor-based industry could result from increased flexibility to set hours and take additional paid holidays or heat exposure on the way to work.[16] The industries with the next largest negative effects are Construction; Transport, Postal and Warehousing; Agriculture, Forestry and Fishing; and Mining, which all have a high proportion of outdoor-based occupations.

Effect sizes for industries that operate essential services like Public Administration and Safety, Education and Training, and Health Care and Social Assistance are relatively small, but still negative (significant at the 10% level). Workers in the Electricity, Gas, Water and Waste services industry seem to have a different effect than others, with work attendance increasing by 0.74 percentage points on hot days ($p$=0.25). Care is needed to not over-interpret this imprecise estimate. Still, a positive effect seems possible for this group, given that electricity networks can be adversely affected on days with extreme temperatures (e.g. due to high use of air conditioning), leading to increased demand for workers (Stone Jr et al., 2021).

To further identify the workers who are most negatively affected by extreme heat, we investigate whether the estimated heat effects (≥38°C) for separate industry-occupation groups can be predicted by observed mean characteristics of the groups. Specifically, we first estimate a heat-effect separately for 77 industry-occupation groups with at least 30,000 observations.[17] The point estimates range from -2.81 (Machinery Operators and Drivers in Construction) to +1.91 (Labourers in Retail), with a mean estimate of -0.78 (see Appendix Table G1). Second, we construct measures of job characteristics for each industry-occupation group using responses from the Household, Income and Labour Dynamics in Australia (HILDA) Survey (see Section

---

[15] Estimates by occupation groups and subgroups based on worker characteristics are provided in Appendix F. The heat-effect is similar across gender, age and parental status. The effect is particularly larger for self-employed workers, possibly due to their flexible working hours or difference in the industry-occupation composition.
[16] Door-to-door sales of insurance policies and financial products is practically non-existent in Australia, and so could not be driving the large negative effect shown in Figure 5.
[17] A large sample is needed due to our within-worker identification approach and the requisite infrequency of extreme heat events.

2 for more details). The measures have been constructed by taking the mean (usual hours, wages and commute time) or proportion of workers with various job characteristics (outdoor, daytime schedule, casual employment, ability to work from home, flexible hours, decide break times and perceived job stress).[18] Third, these characteristics are used as covariates in a regression with the step one estimated ≥38°C coefficient as the dependent variable. The regression is weighted using the number of observations from the Longitudinal Labour Force Survey in each group. Results from this regression are presented in Table 4, illustrating the characteristics that partially explain the variation in heat-estimates across industry-occupation groups.

Column (1) of Table 4 presents estimates using two industry-occupation characteristics as explanatory variables: proportion of workers in the industry-occupation group who work outdoors and the proportion of workers who have daytime work hours. These two characteristics were designed to capture exposure to ambient temperature while at work. The R-squared of this regression implies that these two characteristics explain 20% of the variation in heat estimates between industry-occupations. As discussed throughout the paper, the estimated effects are significantly larger for outdoor-based workers: one standard deviation increase in the proportion of outdoor workers decreases work attendance on days above 38°C by an additional 0.32 ($p<0.01$) percentage points. Jobs with a higher proportion of workers working during the day are also more affected by heat: a one standard deviation increase in the likelihood of daytime work hours decreases work attendance on days over 38°C by 0.29 percentage points ($p<0.01$).

Next, we consider workers with longer commute times, who have greater exposure to heat on their way to work. A one standard deviation increase in commute times is associated with a 0.36 percentage point ($p<0.01$) reduction in the heat-effect. Combined, commute time and the 'heat exposure at work' variables explain 24% of the variation in the heat-estimates between industry-occupations (column 3).

An important factor that might influence the relationship between temperature and work attendance is whether workers are entitled to paid leave (Somanathan et al., 2021).[19] We include a variable that measures industry-occupations with a high proportion of casual employees (column 4). Casual employees in Australia do not have a commitment of ongoing employment or regular work hours, and don't receive paid days off. We also include interaction terms between casual and outdoor and casual and commute time because having access to paid leave may only be important in combination with high heat exposure (i.e. no need to take paid leave on hot days if there is no exposure to heat). Notably, the interaction term between the proportion of casual workers and commute time is associated with a 0.19 reduction in the heat estimate ($p$=0.10). This suggests that the heat-effect for workers with longer commute times is larger in industry-occupations with a greater share of casual workers. A possible explanation is that casual workers take a more short-term transactional perspective when making work attendance decisions (rather than a longer-term career perspective), meaning they are more likely than non-casual workers to be absent if the costs of attending work that day (disutility of work, including during commute) outweigh the benefits from working that day (wages).

---

[18] See Appendix G for a full description of the construction of job characteristics.
[19] Somanathan et al. (2021) distinguish between paid and unpaid leave in their analysis of temperature and absenteeism for Indian manufacturing workers, finding paid leave days increase on hot days for workers with climate control. However, they also report anecdotal evidence of owner-managers of firms citing that workers without paid leave are less willing to work during the summer, suggesting that the "disutility from heat exposure exceeds this difference in wages". Permanent full-time employees in Australia are typically entitled to 4 weeks of annual leave for each 12 months worked.

In Column (5), we control for an additional five factors that may be correlated with other covariates: earnings, the ability to work from home, perceived flexibility to set hours, ability to determine breaks, and perceived job stress. Notably, none of these variables are statistically significant and collectively raise the R-squared by only 5%.[20]

These results help explain the large effect for finance and insurance workers described above. Workers in this industry, on average, have longer average daily commute times (76 mins per day) and 91% work daytime hours, the highest proportion across all industries. On the other hand, workers in the accommodation and food services industry, which is relatively unaffected by heat, spend less time commuting (46 mins per day) and only 40% work a regular daytime schedule. Finance and insurance workers also often commute to work using public transport, spending AU$1,206 annually on public transport and taxis, the highest among all industries (based on an expenditure questionnaire in the HILDA survey).

High temperatures may particularly impact work attendance for workers who rely on public transport for their commute. Extreme heat has been shown to reduce the number of people using public transport (Belloc et al., 2022) and to lower the wellbeing of public transport commuters (Belloc et al., 2023). In addition, extreme temperatures can disrupt transport infrastructure. Train tracks can buckle in high heat, causing delays (Carey, 2013), and Australian train networks reduce speeds on hot days for safety.[21]

---

[20] It is possible that these conditions may be significant in conjunction with heat-exposure (i.e., interaction terms), which we do not rule out. It is also worth noting that the ability to work from home has increased since the onset of COVID-19 pandemic, which occurred subsequent to our analysis period.

Information on commute times and mode are not available in the Australian Longitudinal Labour Force Survey, and so to explore the possible importance of commuting for the temperature-attendance relationship, we use data from the Transport for New South Wales Open Data Platform (Transport for NSW, 2023b). The data allow for the construction of three outcome variables: the number of people entering and exiting each train station per month, the number of bus trips taken in each bus zone per month, and the proportion of trains on each train line that are on time per day. Table 5 presents the estimated effects of temperature on each of these outcomes. Column (1) shows that one additional day exceeding 38°C (instead of a day 26–30°C) reduces train station entries and exits by 1.68% ($p>0.01$). Similarly, monthly bus trips are reduced by 1.28% ($p<0.01$) for each day over 38°C (column 2). These results could be either a consequence or contributing factor of reduced work attendance. In Column (3) of Table 5, we present the effect of temperature on train punctuality and find that days over 38°C have a considerable 11.85 percentage point ($p<0.01$) reduction in on-time service. We don't expect reduced patronage on trains to decrease punctuality; instead, we'd expect the opposite. So, we interpret this result to support our finding that workers with longer commute times, especially those relying on public transport to central business districts of cities, are more sensitive to temperature because of challenges getting to work.

## VII. Is there evidence of adaptation?

Figure 6 shows the effect of daily maximum temperature over 38°C on work attendance for different regions and time periods. Workers and firms may respond differently to temperatures depending on how frequently extreme conditions occur.

To explore regional variation in the effect of heat, we divide our sample into three groups based on the frequency of days over 38°C in the ten years before the analysis period. We find the largest effect in regions in the 1st tercile (historically 0.4–1.2 days per year over 38°C) and the smallest effect in the 3rd (3.3–9.5 days per year). A possible explanation is that regions that experience extreme heat more frequently have invested in long-term adaptation measures, such as cooling infrastructure, partially mitigating the effect of heat. Generally, the return on such investments are larger when the conditions occur more often. On the other hand, another plausible explanation is that the effect is lower in regions with frequent high temperatures because workers are forced to attend work, regardless of adaptation measures. Workers may have financial and work entitlement limits on their ability to be absent from work.

---

[21] We estimate the effect separately for the four largest cities in our sample and find that the heat-effect is largest in Sydney, with work attendance reduced by 2.1 percentage points (p<0.01) on days over 38°C relative to days 26–30°C. Workers in Sydney have longer commute times and a higher proportion take public transport than workers in other areas. Based on the HILDA survey, daily commute times in Sydney average 75 minutes, relative to 73, 68 and 53 minutes for Melbourne, Perth and Adelaide respectively (2011 Greater Capital City Statistical Area). Average annual expenditure on public transportation and taxis is also greater in Sydney ($1,095) than Melbourne ($843), Perth ($539) and Adelaide ($507).

Figure 6 also shows that the heat effect is largest in more recent years. An advantage of our data is that it spans almost 20 years, enabling us to estimate the effect in three groups of 6 years. The most recent year group, between November 2013 and March 2019 has the largest heat-effect. This is also the time when extreme heat has occurred most often.[22] Although we can't rule out other potential explanations, this result may indicate that effects are larger when extreme conditions occur more than the long-run average for the region. In the short term, workers and firms respond to an increase in extreme conditions through reduced work days, but as conditions occur more often, investments in long-term adaptation measures are required.

Graff Zivin & Neidell (2014) find larger, albeit not statistically significant, heat-effects on working hours during the early summer period, which they suggest may indicate acclimatization during the summer. However, we do not find the effect to be different for early months (November and December) compared to late months (February and March); the two-tailed p-value to test for difference in the effect is 0.427.

# VIII. Discussion and Conclusion

Extreme heat significantly reduces both daily work attendance and weekly work hours, with non-attendance accounting for approximately 69% of the overall reduction, while the remaining 31% is attributed to shorter workdays. Notably, the effects are concentrated on the same day as the extreme heat, and lost work time is not compensated for in the surrounding days and weeks. Therefore, reduced work time may explain heat-related output losses, in addition to the possibility of less efficient work on hot days.

---

[22] The number of days over 38°C increased throughout the analysis period for all SA4 regions. However, the distrubtion of temperaures for days over 38°C remained stable. The mean maximum temperature for days reaching over 38°C was 40.0°C, 40.2°C and 40.2°C for the three year groups respectively. Thus, we do not attribute the difference in effect between time periods to a difference in the treatment intensity.

Our results are based on over 9 million days from 649,740 Australian workers enabling estimation of industry-specific heat-effects. We find considerable heterogeneity in the effect of temperature across the labor force. Importantly, coarsely categorizing workers into indoor- and outdoor-based jobs inadequately explains which industries are most temperature-sensitive. Our results suggest that firms and workers alter work time for multiple reasons. In particular, heat exposure during the work commute may be a barrier for workers to attend work on extreme heat days. This conclusion is supported by our finding of a significant decrease in public transport patronage and an increase in delayed services on days over 38 °C.

Our results differ from those in Graff Zivin & Neidell (2014). They find small and statistically insignificant effects for their total sample and sample of people in low-risk industries, concluding that "time allocated to labor on net is not responsive to changes in temperature." Our results also differ from Dillender (2021), who find no statistically significant effects on hours worked by outdoor and indoor workers (using their full sample of regions). Results for outdoor-based workers are more consistent, with Graff Zivin & Neidell (2014) also finding a large statistically significant effect when temperatures are over 100°F.

The literature has typically focused on labour supply mechanisms to explain the strong negative relationship between extreme heat and work hours.[23] However, our results could also be explained by demand-side factors. Some workplaces have heat thresholds for the cessation of work to protect workers from working in unsafe environments. Similarly, plant breakdown and insufficient demand on high-heat days could lead to workers being stood down. Further research is needed to understand the extent to which our findings can be (partially) explained by changes in labour demand on hot days.

Reduced work time on extreme heat days should not necessarily be interpreted as undesirable. In fact, it implies that fewer workers are exposed to riskier conditions (Dillender, 2021; Ireland et al., 2023) and less productive workdays (LoPalo, 2023; Seppanen et al., 2006). The policy goal should not be to eliminate the temperature-labor relationship, but rather to enable firms and workers to make informed decisions for efficient resource allocation. Allowing workers the flexibility to adjust their leave in response to extreme heat forecasts could result in net gains in both welfare and output, particularly if work is reallocated to more productive and safer conditions.

---

[23] Lai et al. (2023) note in their review that "the literature discussing the effect of temperature on firms' labor demand is very limited" (p.216).

# References

Barreca, A., Clay, K., Deschenes, O., Greenstone, M., & Shapiro, J.S. (2016). Adapting to climate change: The remarkable decline in the US temperature-mortality relationship over the twentieth century. *Journal of Political Economy,* 124, 105-159.

Behrer, A.P., & Park, J. (2017). Will we adapt? temperature, labor and adaptation to climate change. *Harvard Project on Climate Agreements Working Paper*, 16-81.

Belloc, I., Gimenez-Nadal, J.I., & Molina, J.A. (2022). Weather conditions and daily commuting. *IZA Discussion Papers,* No. 15661.

Belloc, I., Gimenez-Nadal, J.I., & Molina, J.A. (2023). Feelings in Travel Episodes and Extreme Temperatures. *IZA Discussion Papers,* No. 16241.

Bureau of Meteorology. (2022). Climate classification maps. Date Accessed: 2022-06-15 http://www.bom.gov.au/jsp/ncc/climate_averages/climate-classifications/index.jsp?maptype=kpn#maps

Carey, A. (2013). Why our rails can't cope with the heat. The Age.

Carias, M.E., Johnston, D., Knott, R., & Sweeney, R. (2022). The Effects of Temperature on Economic Preferences. *arXiv preprint arXiv:2110.05611.*

CSIRO and Bureau of Meteorology. (2022). Climate Change in Australia. Date Accessed: 11/2/2022 http://www.climatechangeinaustralia.gov.au

Deryugina, T., & Hsiang, S.M. (2014). Does the environment still matter? Daily temperature and income in the United States. National Bureau of Economic Research.

Dillender, M. (2021). Climate Change and Occupational Health Are There Limits to Our Ability to Adapt? *Journal of Human Resources,* 56, 184-224.

Freeman, R.B. (1997). Working for nothing: The supply of volunteer labor. *Journal of Labor Economics,* 15, S140-S166.

Frey, B.S., & Jegen, R. (2001). Motivation crowding theory. *Journal of economic surveys,* 15, 589-611.

Garg, T., Gibson, M., & Sun, F. (2020). Extreme temperatures and time use in China. *Journal of Economic Behavior & Organization,* 180, 309-324.

Global Modeling Assimilation Office. (2015). MERRA-2 inst3_3d_asm_Np: 3d, 3-hourly, instantaneous, pressure-level, assimilation, assimilated meteorological fields V5. 12.4. In M. Greenbelt, USA, Goddard Earth Sciences Data and Information Services Center (GES DISC) (Ed.).

Graff Zivin, J., & Neidell, M. (2014). Temperature and the allocation of time: Implications for climate change. *Journal of Labor Economics,* 32, 1-26.

Hansen, J., & Lebedeff, S. (1987). Global trends of measured surface air temperature. *Journal of geophysical research: Atmospheres,* 92, 13345-13372.

Heyes, A., & Saberian, S. (2022). Hot Days, the ability to Work and climate resilience: Evidence from a representative sample of 42,152 Indian households. *Journal of Development Economics,* 155, 102786.

Ireland, A., Johnston, D., & Knott, R. (2023). Heat and worker health. *Journal of health economics,* 91.

Johnston, D.W., Knott, R., Mendolia, S., & Siminski, P. (2021). Upside-Down Down-Under: Cold Temperatures Reduce Learning in Australia. *Economics of Education Review,* 85, 102172.

Lai, W., Qiu, Y., Tang, Q., Xi, C., & Zhang, P. (2023). The Effects of Temperature on Labor Productivity. *Annual Review of Resource Economics,* 15.

Li, W. (2020). Is Working from Home a Way of Adaptation to Climate Change?

LoPalo, M. (2023). Temperature, worker productivity, and adaptation: evidence from survey data production. *American Economic Journal: Applied Economics,* 15, 192-229.

Metro Trains. (2023). Hot Weather and Speed Restrictions. Date Accessed: 2023-08-14 https://www.metrotrains.com.au/hot-weather/

National Center for O*NET Development. (2021). Work Context: Outdoors, Exposed to Weather. *O*NET OnLine*. Date Accessed: 2/11/2021 www.onetonline.org/find/descriptor/result/4.C.2.a.1.c?a=1

Neidell, M., Graff Zivin, J., Sheahan, M., Willwerth, J., Fant, C., Sarofim, M., et al. (2021). Temperature and work: Time allocated to work under varying climate and labor market conditions. *PloS one,* 16, e0254224.

NOAA National Weather Service. (2023). What is the heat index. Date Accessed: 2023-06-27 https://www.weather.gov/ama/heatindex

Public Transport Authority. (2019). Potential for extreme heat to slow train services. Date Accessed: 2023-08-14 https://www.pta.wa.gov.au/news/media-statements/potential-for-extreme-heat-to-slow-train-services-1

Rode, A., Baker, R.E., Tamma Carleton, A.L., D'Agostino, M.D., Foreman, T., Gergel, D.R., et al. (2023). Is workplace temperature a valuable job amenity? Implications for climate change.

Sahu, S., Sett, M., & Kjellstrom, T. (2013). Heat exposure, cardiovascular stress and work productivity in rice harvesters in India: implications for a climate change future. *Ind Health,* 51, 424-431.

Seppanen, O., Fisk, W.J., & Lei, Q. (2006). Effect of temperature on task performance in office environment. *Lawrence Berkeley National Lab. Report Number LBNL-60946*.

Somanathan, E., Somanathan, R., Sudarshan, A., & Tewari, M. (2021). The impact of temperature on productivity and labor supply: Evidence from Indian manufacturing. *Journal of Political Economy,* 129, 1797-1827.

Stone Jr, B., Mallen, E., Rajput, M., Broadbent, A., Krayenhoff, E.S., Augenbroe, G., et al. (2021). Climate change and infrastructure risk: Indoor heat exposure during a concurrent heat wave and blackout event in Phoenix, Arizona. *Urban Climate,* 36, 100787.

Transport for NSW. (2023a). Train speed restrictions during hot weather. Date Accessed: 2023-08-14 https://transportnsw.info/travel-info/ways-to-get-around/train/train-speed-restrictions-during-hot-weather

Transport for NSW. (2023b). Open Data Hub. Date Accessed: 2023-07-17 https://opendata.transport.nsw.gov.au/

Wooden, M., & Watson, N. (2007). The HILDA survey and its contribution to economic and social research (so far). *Economic record,* 83, 208-231.

Zhang, P., Deschenes, O., Meng, K., & Zhang, J. (2018). Temperature effects on productivity and factor reallocation: Evidence from a half million Chinese manufacturing plants. *Journal of Environmental Economics and Management,* 88, 1-17.

# Tables & Figures

**Table 1. Descriptive Statistics from Longitudinal Labour Force Survey**

| | |
|---|---|
| **Day level panel, attended work** | |
| Number of responses, N (%) | 9,140,637 (100) |
|    - Attended, N (%) | 7,420,799 (81.2) |
|    - Absent, N (%) | 1,719,838 (18.8) |
| | |
| **Week level panel, hours worked** | |
| Number of responses, N | 1,688,074 (100) |
|    - Actual hours per week, µ (SD) | 35.2 (15.3) |
|    - Usual hours per week, µ (SD) | 36.0 (14.3) |
|    - Weekdays worked per week, µ (SD) | 4.1 (1.3) |
|    - Total days worked per week, µ (SD) | 4.5 (1.3) |
| | |
| **Number of individuals, N** | 649,740 |
|    - Male, N (%) | 350,886 (54.0) |
|    - Age, µ (SD) | 39.6 (13.6) |
| | |
| **Number of households, N** | 358,232 |

**Notes:** Based on single-job holders November–March from November 2001–February 2020 for included SA4 regions. Day-of-week restricted to days the individual ever reported to attend work (i.e. Sundays only included for individuals who ever worked on a Sunday). Each response contains attendance and working hours for a reference week (i.e. 7 days of attendance responses per monthly response).

**Table 2. Effect of Temperature on Hours Worked**

|  | (1) Hours per Week | (2) Days per Week | (3) Hours per Day | (4) Less than usual hours |
|---|---|---|---|---|
| *Panel A. All Workers (N=1,514,977)* | | | | |
| 34–38°C | -0.26*** | -0.03*** | -0.01 | 2.04*** |
|  | (0.07) | (0.01) | (0.01) | (0.59) |
| ≥38°C | -0.77*** | -0.06*** | -0.05** | 6.47*** |
|  | (0.14) | (0.02) | (0.02) | (1.10) |
| Outcome Mean | 35.54 | 4.50 | 7.75 | 21.64 |
| *Panel B. Outdoor Workers (N=135,002)* | | | | |
| 34–38°C | -0.49*** | -0.04** | -0.06** | 4.33*** |
|  | (0.18) | (0.02) | (0.03) | (1.00) |
| ≥38°C | -2.02*** | -0.13** | -0.13 | 14.14*** |
|  | (0.58) | (0.05) | (0.11) | (2.61) |
| Outcome Mean | 40.28 | 4.83 | 8.30 | 23.77 |
| *Panel C. Indoor Workers (N=1,342,350)* | | | | |
| 34–38°C | -0.22*** | -0.02*** | -0.00 | 1.72*** |
|  | (0.07) | (0.01) | (0.01) | (0.58) |
| ≥38°C | -0.71*** | -0.06*** | -0.04 | 5.88*** |
|  | (0.12) | (0.02) | (0.03) | (1.16) |
| Outcome Mean | 35.30 | 4.48 | 7.72 | 21.48 |

**Notes:** Based on linear regression specification includes bins for average daily maximum temperature, individual-area, month-year and week-of-year fixed effects and controls for precipitation and public holidays. Outcome variables are weekly hours worked in Column (1); days worked per week in Column (2); hours worked per day in Column (3); and a binary outcome indicating working less than usual hours in Column (4). Average daily maximum temperature compared to reference category of 26–30°C. Sample includes single job holders for the period November–March from November 2001–February 2020. Sample size and outcome mean excludes singletons. Standard errors clustered by area. ***p<0.01 **p<0.05 *p<0.1

**Table 3. Effect of Heat on Work Attendance, Previous Temperature**

|  | (1) Today | (2) Yesterday | (3) Midnight | (4) All |
|---|---|---|---|---|
| *Panel A. All Workers (N=9,129,807 μ=81.17)* | | | | |
| Today ≥ 38°C | -0.95*** (0.12) | | | -0.88*** (0.12) |
| Yesterday ≥ 38°C | | -0.45*** (0.14) | | -0.27* (0.15) |
| Midnight ≥ 26°C | | | -0.36*** (0.12) | -0.09 (0.14) |
| *Panel B. Outdoor Workers (N=856,050 μ=81.36)* | | | | |
| Today ≥ 38°C | -1.67*** (0.24) | | | -1.52*** (0.24) |
| Yesterday ≥ 38°C | | -0.86*** (0.28) | | -0.53* (0.26) |
| Midnight ≥ 26°C | | | -0.73** (0.27) | -0.22 (0.25) |
| *Panel C. Indoor Workers (N=7,998,058 μ=81.13)* | | | | |
| Today ≥ 38°C | -0.83*** (0.13) | | | -0.77*** (0.13) |
| Yesterday ≥ 38°C | | -0.41*** (0.15) | | -0.26 (0.17) |
| Midnight ≥ 26°C | | | -0.32** (0.14) | -0.07 (0.16) |

**Notes:** Based on linear regression with binary outcome for daily work attendance (zero to indicate absence, 100 for attendance). Regression specification includes bins for daily maximum temperature, individual-area, month-year and day-of-year fixed effects and controls for day-of-week and precipitation. Indicator variables for maximum temperature greater than 38°C and midnight temperature over 26°C. Outdoor and Indoor classification based upon occupation codes. For the period November–March from November 2001–February 2020. Sample size and outcome mean excludes singletons. Standard errors clustered by area. ***p<0.01 **p<0.05 *p<0.1

**Table 4. Importance of work characteristics on the effect of extreme heat on work attendance**

|  | (1) | (2) | (3) | (4) | (5) |
|---|---|---|---|---|---|
| Outdoor | -0.32*** | - | -0.25** | -0.15 | -0.06 |
|  | (0.10) |  | (0.10) | (0.11) | (0.14) |
| Daytime schedule | -0.29*** | - | -0.18* | -0.10 | -0.07 |
|  | (0.09) |  | (0.11) | (0.13) | (0.14) |
| Commute time | - | -0.36*** | -0.21* | -0.22* | -0.24* |
|  |  | (0.10) | (0.11) | (0.11) | (0.13) |
| Casual (no paid leave) | - | - | - | -0.06 | -0.20 |
|  |  |  |  | (0.12) | (0.22) |
| Casual (no paid leave) × outdoor | - | - | - | -0.10 | -0.10 |
|  |  |  |  | (0.13) | (0.14) |
| Casual (no paid leave) × commute time | - | - | - | -0.19* | -0.26** |
|  |  |  |  | (0.11) | (0.13) |
| Controls (work conditions) | - | - | - | - | Yes |
| R-squared | 0.20 | 0.15 | 0.24 | 0.29 | 0.34 |

**Notes:** Based on largest industry-occupation groups (N=77). Linear regression with the estimated heat-effect as the dependent variable (over 38°C on work attendance), weighted by group sample size (from Labour Force Survey). Outcome mean is -0.78. Job characteristics for industry-occupation groups are based on O*NET and responses to the HILDA survey (see Appendix G for details). All regressors are standardized by demeaning and dividing by the standard deviation for comparison. Standard errors in parentheses. Column (5) includes controls for earnings, ability to work from home, flexible hours, decide break times and perceived job stress. Coefficients of controls variables reported in Appendix G, Table G2.  ***p<0.01 **p<0.05 *p<0.1

**Table 5. Effect of Temperature on Public Transport Usage and Performance**

|  | (1) Ln(Train Station Entries and Exits) | (2) Ln(Bus Trips) | (3) Train Punctuality |
|---|---|---|---|
| <22 | 0.30 | -0.22 | -5.34*** |
|  | (0.34) | (0.60) | (1.59) |
| 22-26 | -0.12 | 0.33* | -0.76 |
|  | (0.12) | (0.18) | (0.55) |
| 26-30 | - | - | - |
| 30-34 | -0.08 | -0.22 | -1.39** |
|  | (0.15) | (0.32) | (0.57) |
| 34-38 | -0.19 | -0.26 | -3.03*** |
|  | (0.20) | (0.26) | (0.96) |
| ≥38 | -1.68*** | -1.28*** | -11.85*** |
|  | (0.26) | (0.43) | (1.86) |
| Level | Train station (N=314) | Bus Zone (N=26) | Train Line (N=17) |
| Frequency | Monthly | Monthly | Daily |
| Fixed Effects | Station-Year; Station-Month | Zone-Year; Zone-Month | Line-Year-Month |
| Period | Nov2016–Feb2019 | Nov2016–Feb2019 | Nov2018–Feb2019 |
| Standard Errors | Robust, Clustered by Train Station | Robust, Clustered by Bus Zone | Robust |
| Outcome Mean (untransformed) | 208,821 | 939,786 | 88.95 |
| N | 5,814 (Station-Months) | 475 (Zone-Months) | 6,065 (Line-Days) |

**Notes:** Linear Regression for November-March only. Outcome variables are (i) log of monthly train station entry and exits; (ii) log of monthly bus trips and; (iii) daily train punctuality (proportion of on-time trips). Estimates are multiplied by 100 for interpretation. Standard errors below coefficients in parentheses. Coefficients shows the effect of an additional day during the month with maximum temperature in bins, relative to 26–30°C. See Appendix H for further details. ***p<0.01 **p<0.05 *p<0.1

**Figure 1. Effect of Daily Maximum Temperature on Work Attendance**

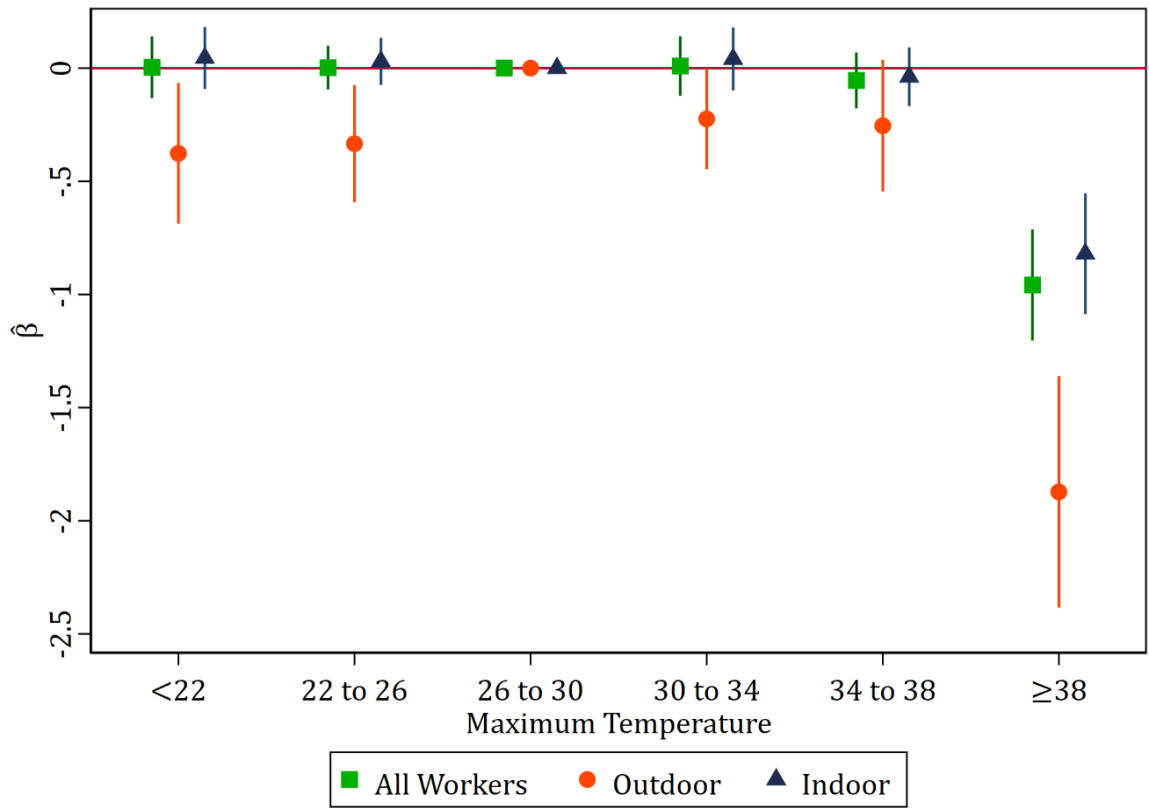

**Notes:** Based on linear regression with a rescaled binary outcome for daily work attendance (zero to indicate absence, 100 for attendance). Sample includes single job holders for the period November–March from November 2001–February 2020. Regression specification includes bins for daily maximum temperature, individual-area, month-year and calendar day-of-year fixed effects and controls for day-of-week and precipitation. 95% confidence intervals, standard errors clustered by area. Outdoor and Indoor classification based on occupation codes.

**Figure 2. Effect of Average Daily Maximum Temperature on Hours Worked**

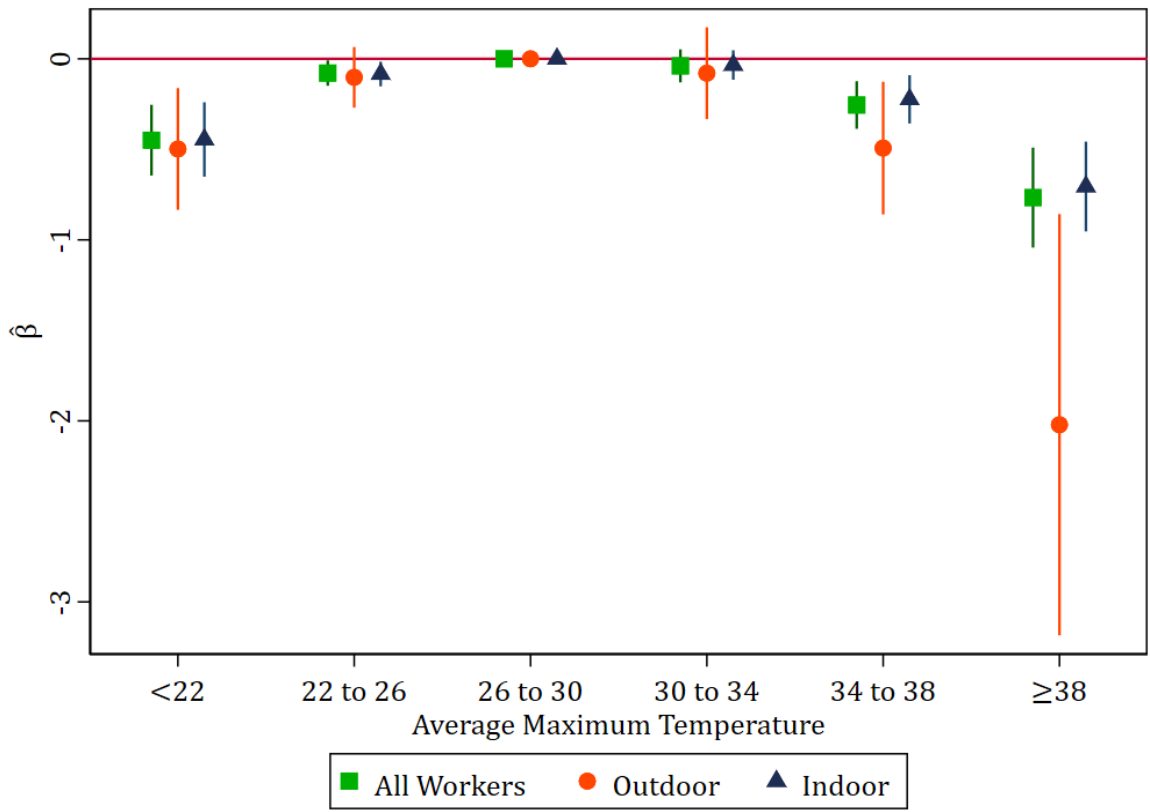

**Notes:** Based on linear regression with weekly hours worked as outcome. Regression specification includes bins for average daily maximum temperature, individual-area, month-year and week-of-year fixed effects and controls for precipitation and public holidays. Average maximum temperature, denotes the average of daily maximum temperatures throughout the week. Sample includes single job holders for the period November–March from November 2001–February 2020. 95% confidence intervals, standard errors clustered by area. Outdoor and Indoor classification based on occupation codes.

**Figure 3. Effect of Temperature on Working Less than Usual Hours by Reason**

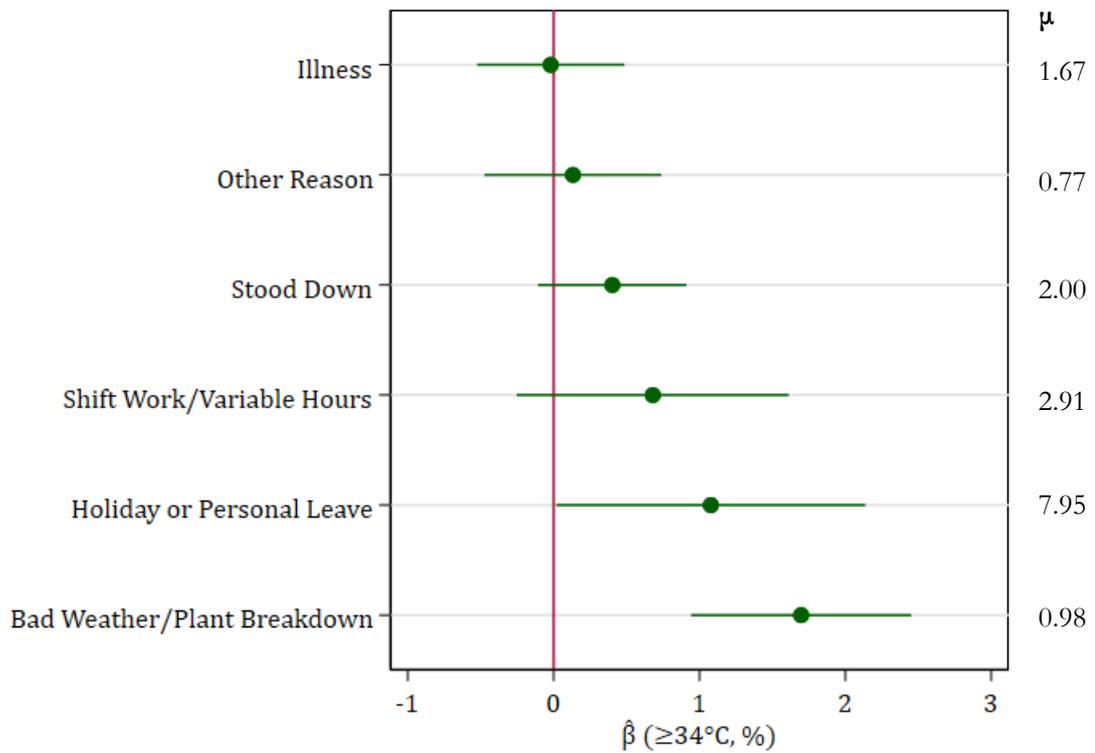

**Notes:** Based on linear regression with binary outcome indicating working less than usual hours during the week for each reason. Outdoor workers only, based on occupation (see Appendix D, Figure D2 for indoor workers). Regression specification includes average daily maximum temperature ≥34°C as an indicator variable, individual-area, month-year and week-of-year fixed effects and controls for precipitation and public holidays. 95% confidence intervals, standard errors clustered by area. Outcome mean (μ) displayed on the right of chart. Sample includes single job holders for the period November–March from November 2001–February 2020.

**Figure 4. Effect of Extreme Temperature on Hours Worked, Past and Prior Weeks**

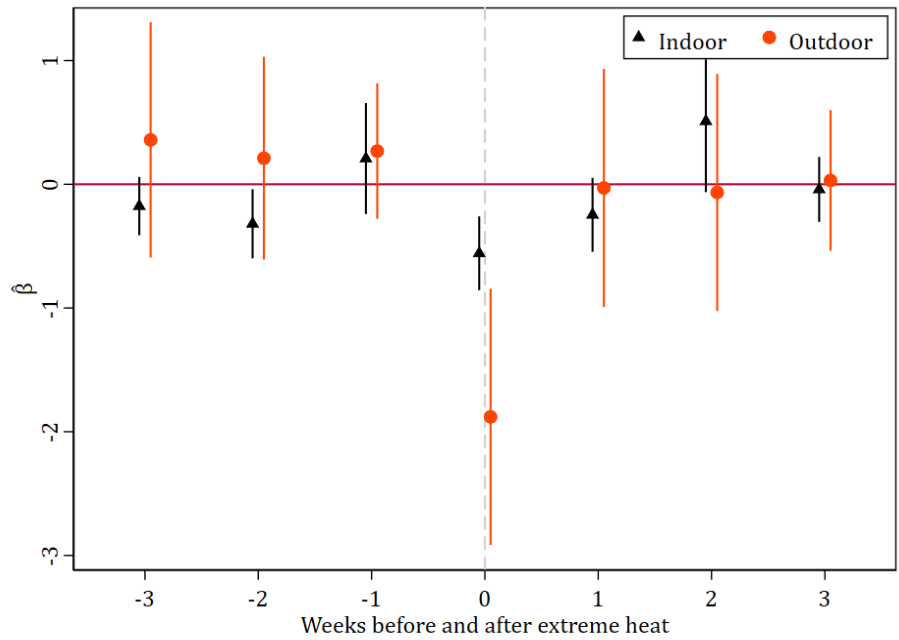

**Notes:** Based on linear regression with weekly hours worked as outcome. The specification includes seven indicator variables for average daily maximum temperature ≥38°C in current week and the three weeks before and after. Regression specification also includes individual-area, month-year and week-of-year fixed effects and controls for precipitation and public holidays. Sample includes single job holders for the period November–March from November 2001–February 2020. 95% confidence intervals, standard errors clustered by area. Outdoor and Indoor classification based on occupation codes.

**Figure 5. Effect of Temperature Exceeding 38°C on Work Attendance by Industry**

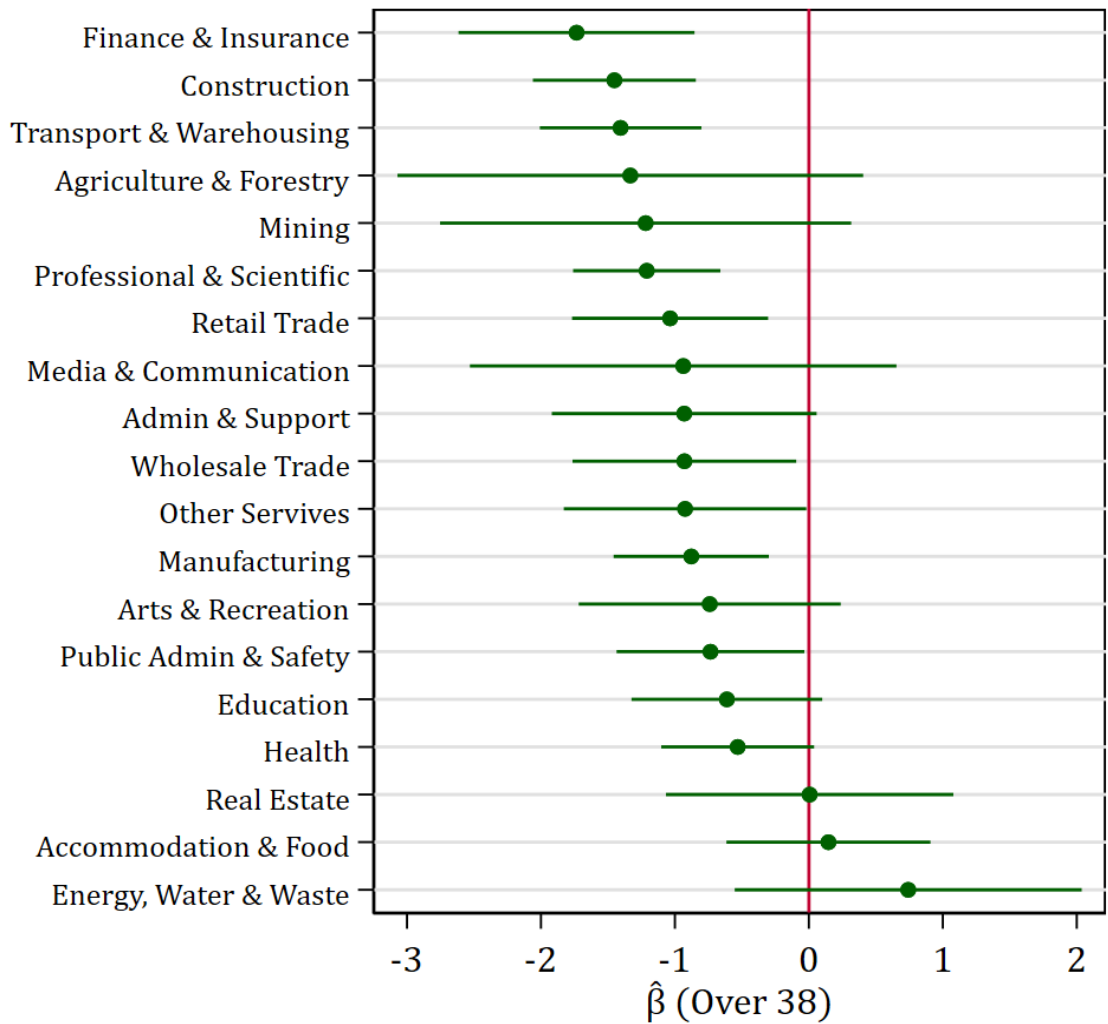

**Notes:** Based on linear regression with binary outcome for daily work attendance (zero to indicate absence, 100 for attendance). Regression specification includes bins for daily maximum temperature, individual-area, month-year and day-of-year fixed effects and controls for day-of-week and precipitation. 95% confidence intervals, standard errors clustered by area. Over 38°C compared to reference category of 26–30°C. Industry based on ANZSIC 2006 (descriptions have been abbreviated). Sample includes single job holders for the period November–March from November 2001–February 2020.

**Figure 6. Effect of Temperature Exceeding 38°C on Work Attendance by Region and Year**

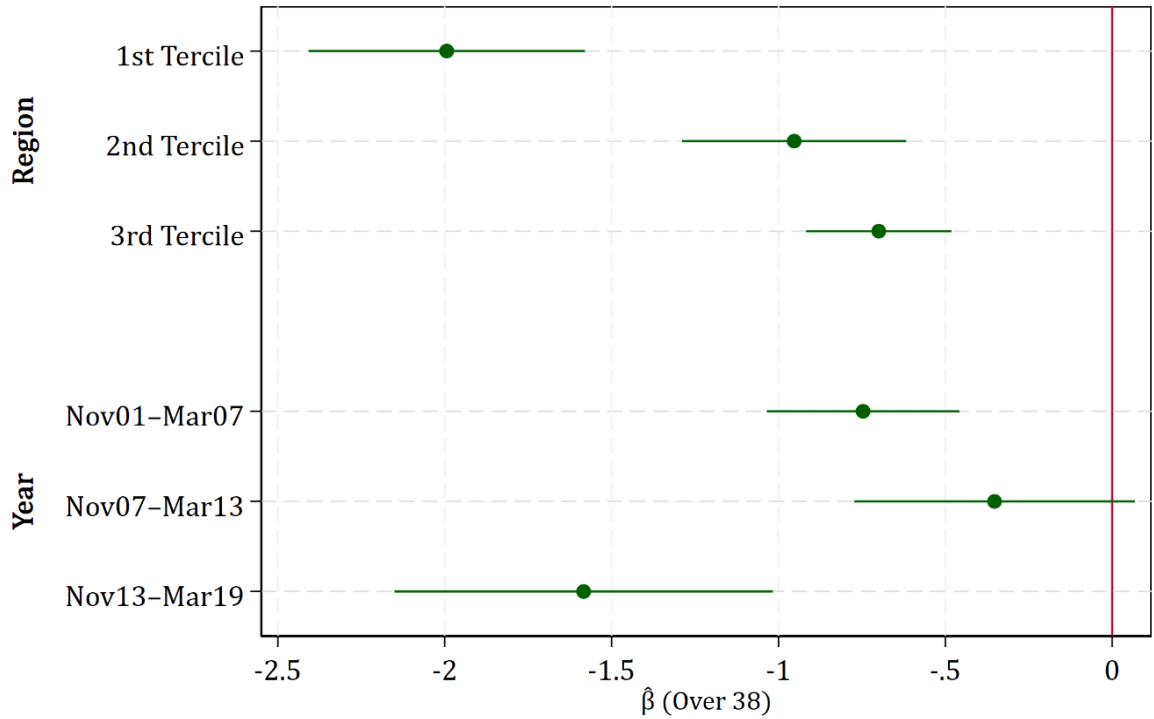

**Notes:** Based on linear regression with binary outcome for daily work attendance (zero to indicate absence, 100 for attendance). Regression specification includes bins for daily maximum temperature, individual-area, month-year and day-of-year fixed effects and controls for day-of-week and precipitation. Over 38°C compared to reference category of 26–30°C. Sample includes single job holders for the period November–March from November 2001–February 2020. Climate terciles based on frequency of days exceeding 38°C from least (1st tercile) to most (3rd tercile) in the 10 years prior to analysis period (1991–2000). 95% confidence intervals, robust standard errors for climate regressions and robust standard errors clustered by area for year regressions.

# Appendices

**Appendix A. Inclusion Criteria for Statistical Areas**

The included areas are shown below in Figure A1. We exclude area in climates that rarely experience extreme heat days. Specifically, we focus on warm and mild temperate climates (between 30–40°S). We also exclude area that are too large for climatic variables at single point to be sufficiently representative for workers in that area. SA4 (2016) regions vary greatly in area from 58km$^2$ for Sydney's Eastern Suburbs to 1,372,033km$^2$ covering the southern outback of Western Australia. Large SA4 regions are can have multiple distant urban areas. An example of an excluded area is North West Victoria (Figure A2), which is 78,000km$^2$ and does not have a single central population center. The two major labor markets in the area, Horsham and Mildura, are 281km apart and experience very different climates.

**Figure A1. Map of Included Regions**

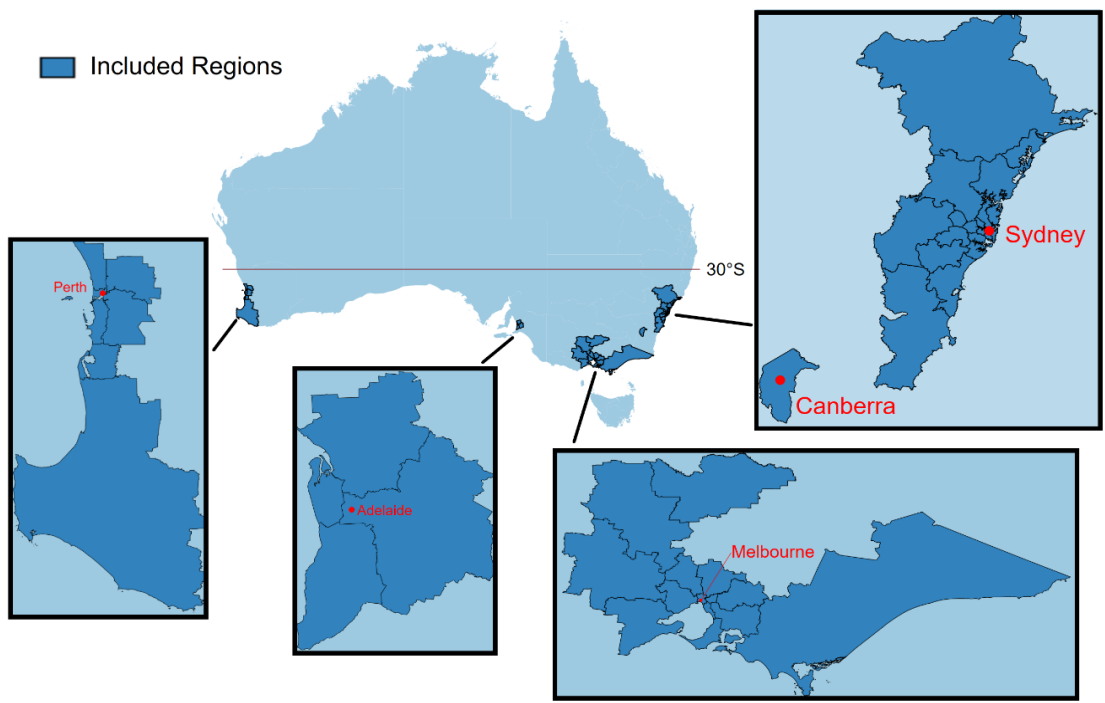

**Notes:** Map of Australia with the 45 regions, SA4, included in this study shaded. All regions are between 30°S–40°S.

**Figure A2. Example of Statistical Area Level 4 - North West Victoria**

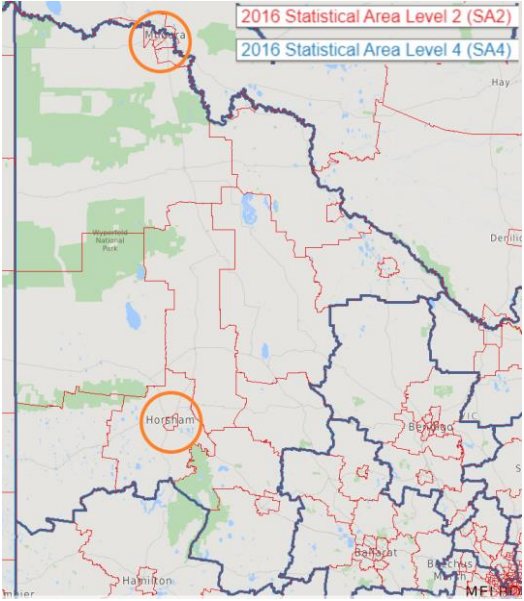

**Notes:** Map of North West Victoria, an example of a large region that did not meet the inclusion criteria. The area has two major urban areas (highlighted in orange) with different climates.

# Appendix B. Descriptive Statistics from Longitudinal Labour Force Survey

**Figure B1. Attendance Rate and Responses by Calendar Date**

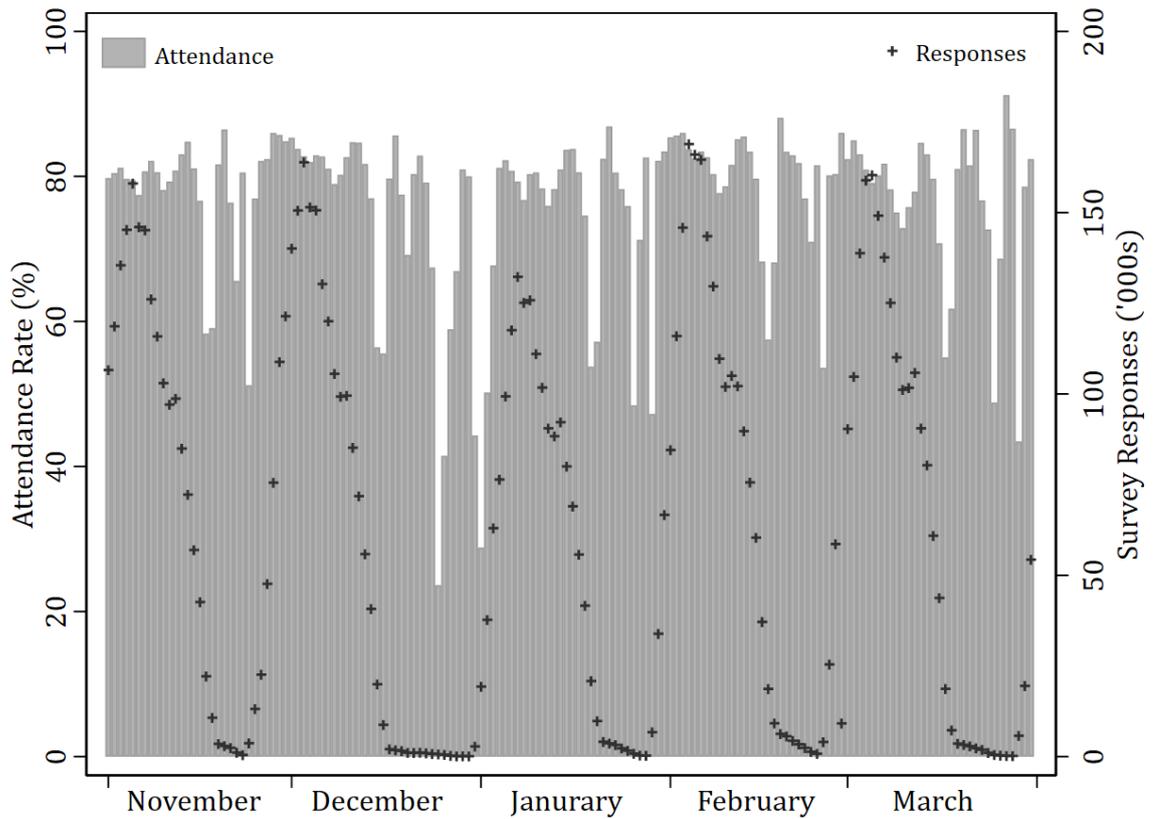

**Notes:** Based on single-job holders November–March from November 2001-February 2020 for included SA4 regions. Values represent the average over the 19-year period. Restricted to individuals who worked at least one day during the week and day of week is restricted for individuals to days they ever reported attending work, across any month (i.e. responses on Sundays are only included for individuals who ever attended work on a Sunday).

**Table B1. Descriptive Statistics from Longitudinal Labour Force Survey**

|  | Responses, week level | | Average Attendance (%) | Average weekly hours worked |
|---|---|---|---|---|
|  | N | % |  |  |
| Male | 923,332 | 54.7 | 83.2 | 39.2 (14.9) |
| Female | 764,742 | 45.3 | 78.7 | 30.3 (14.2) |
|  |  |  |  |  |
| Age 15-17 | 48,150 | 2.9 | 56.6 | 14.8 (12.1) |
| Age 18-29 | 411,897 | 24.4 | 79.0 | 33.6 (14.0) |
| Age 30-49 | 783,070 | 46.4 | 83.0 | 37.2 (14.7) |
| Age 50-64 | 397,822 | 23.6 | 82.9 | 36.4 (15.4) |
| Age 65+ | 47,135 | 2.8 | 77.6 | 27.7 (17.9) |
|  |  |  |  |  |
| Outdoor | 149,718 | 9.2 | 81.4 | 40.0 (15.4) |
| Indoor | 1,470,852 | 90.8 | 81.1 | 35.0 (15.0) |
| All | 1,688,074 | 100 | 81.2 | 35.2 (15.3) |

**Notes:** Based on single-job holders November–March from November 2001-February 2020 for included SA4 regions. Day-of-week restricted to days the individual ever reported to attend work (i.e. Sundays only included for individuals who ever worked on a Sunday). Each response contains attendance and working hours for a reference week (i.e. 7 days of attendance responses per monthly response). Outdoor and Indoor classification based on occupation (unavailable for 1.0% of responses).

**Figure B2. Attendance Rate by Day-of-week**

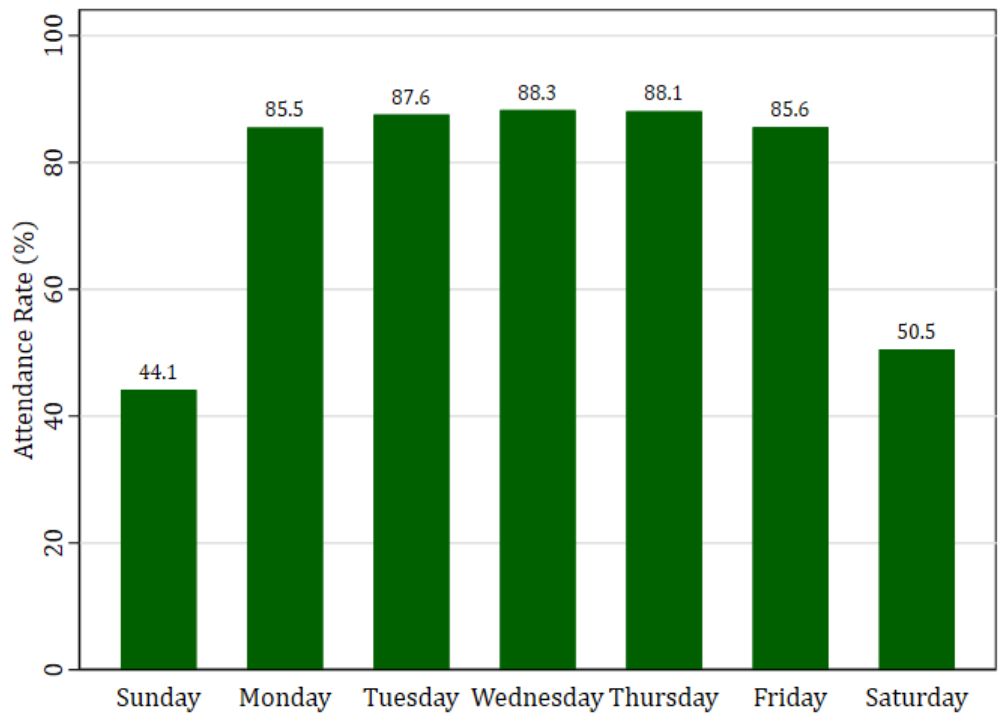

**Notes:** Based on single-job holders November–March from November 2001-February 2020 for included SA4 regions. Day-of-week restricted to days the individual ever reported to attend work (i.e. Sundays only included for individuals who ever worked on a Sunday).

**Appendix C. Climate Classification of Australia**

**Table C1. Annual heat days (≥34°C) and extreme heat days (≥38°C) for Australian capital cities**

|  |  | Actual | Intermediate Projection | | | High Projection | | |
|---|---|---|---|---|---|---|---|---|
|  |  | 1981–2010 | 2030 | 2050 | 2070 | 2030 | 2050 | 2070 |
| Sydney | 34°C | 7.4 | 10.6 | 12.5 | 14.2 | 11.2 | 15.1 | 19.6 |
|  | 38°C | 1.2 | 2.0 | 2.5 | 3.0 | 2.2 | 3.3 | 4.8 |
| Melbourne | 34°C | 11.5 | 14.7 | 16.4 | 18.4 | 15.0 | 18.7 | 22.9 |
|  | 38°C | 2.7 | 4.1 | 4.9 | 6.0 | 4.3 | 6.0 | 8.1 |
| Brisbane* | 34°C | 3.8 | 7.8 | 11.1 | 12.9 | 9.6 | 16.2 | 27.5 |
|  | 38°C | 0.3 | 0.4 | 0.6 | 0.8 | 0.5 | 0.9 | 1.7 |
| Perth | 34°C | 22.9 | 31.0 | 34.4 | 38.5 | 30.1 | 39.2 | 47.6 |
|  | 38°C | 5.2 | 8.2 | 9.5 | 11.4 | 8.0 | 11.5 | 15.6 |
| Adelaide | 34°C | 22.1 | 27.4 | 29.3 | 32.0 | 28.0 | 32.6 | 38.2 |
|  | 38°C | 6.3 | 9.6 | 10.7 | 12.2 | 9.9 | 12.6 | 15.7 |
| Darwin* | 34°C | 119.7 | 192.1 | 222.1 | 246.5 | 203.1 | 251.3 | 300.1 |
|  | 38°C | 0.4 | 1.5 | 3.0 | 5.2 | 1.8 | 7.3 | 28.5 |
| Canberra | 34°C | 11.1 | 16.1 | 18.6 | 21.9 | 16.9 | 22.8 | 28.2 |
|  | 38°C | 1.1 | 2.6 | 3.2 | 4.5 | 2.8 | 4.8 | 7.1 |
| Hobart* | 34°C | 1.1 | 1.5 | 1.9 | 2.1 | 1.7 | 2.1 | 2.9 |
|  | 38°C | 0.0 | 0.1 | 0.3 | 0.3 | 0.2 | 0.3 | 0.6 |

**Notes:** Data sourced from climate change in Australia website (CSIRO and Bureau of Meteorology, 2022). Forecasts for 2030 (2016-2045), 2050 (2036-2065 and 2070 (2056-2085) under intermediate (RCP 4.5) and high (RCP 8.5) greenhouse gas concentration scenarios. Mean of estimates from 8 climate models (ACCESS1-0, CESM1-CAM5, CNRM-CM5, CanESM2, GFDL-ESM2M, HadGEM2-CC, MIROC5 and NorESM1-M). *Not included in this study.

**Figure C1. Annual Days over 38°C (100°F) in Included Regions**

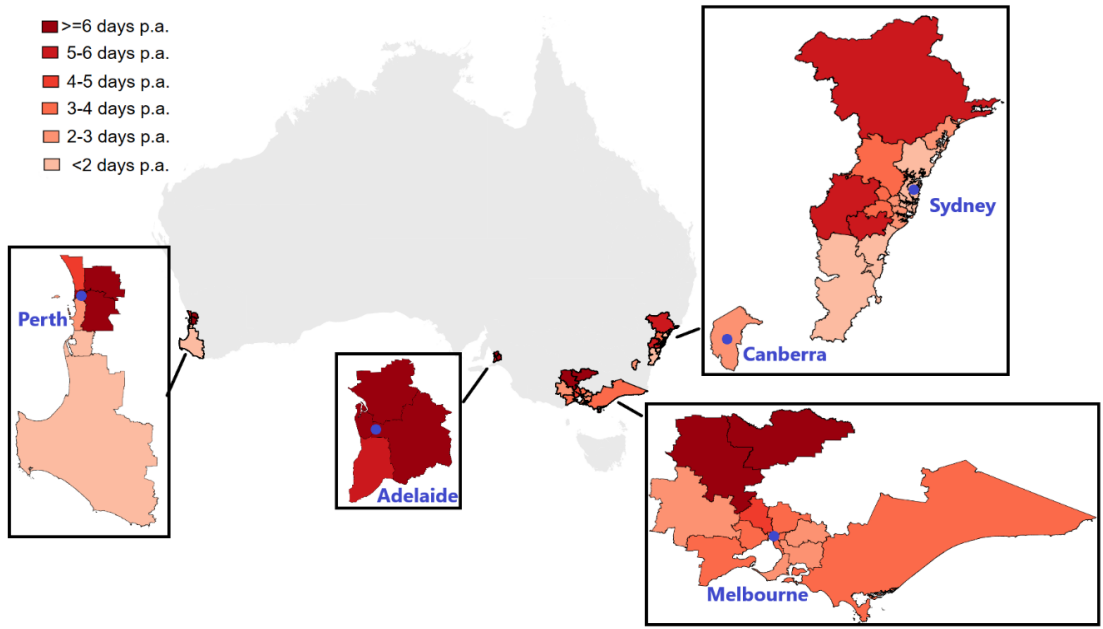

**Notes:** Figures shows included SA4 regions. Based on November 2001 to February 2020.

**Figure C2. Climate Classification of Australia**

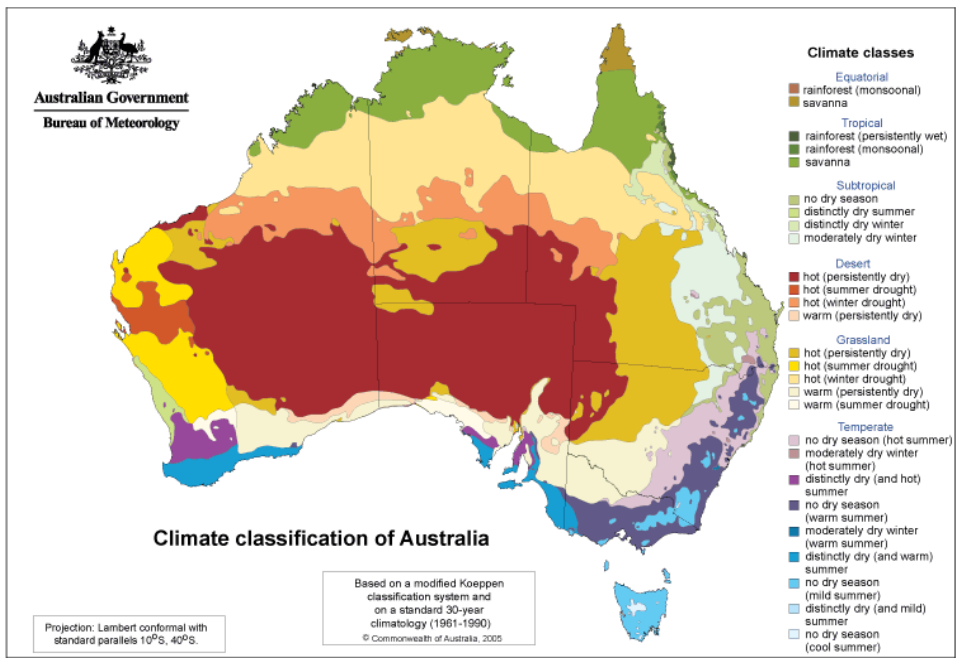

**Notes:** Based on a modified Köppen climate classification system (Bureau of Meteorology, 2022).

**Figure C3. Distribution of maximum temperature**

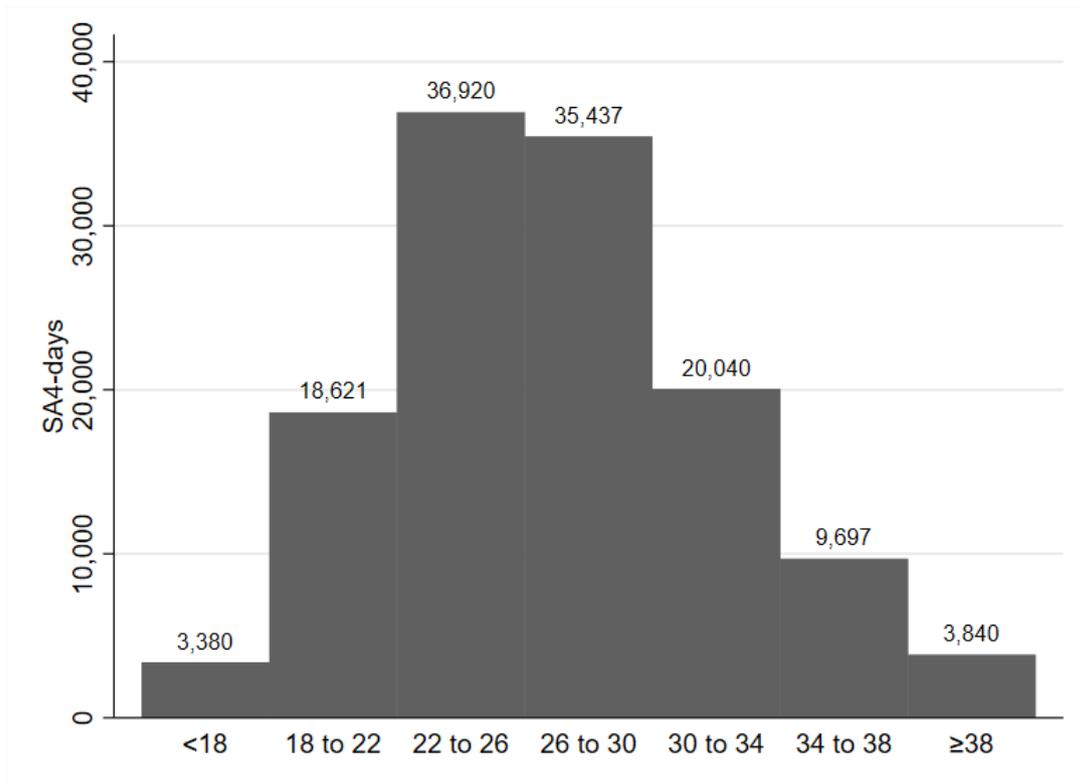

**Notes:** Based on November–March from November 2001–February 2020 for included SA4 regions.

**Figure C4. Distribution of maximum temperature**

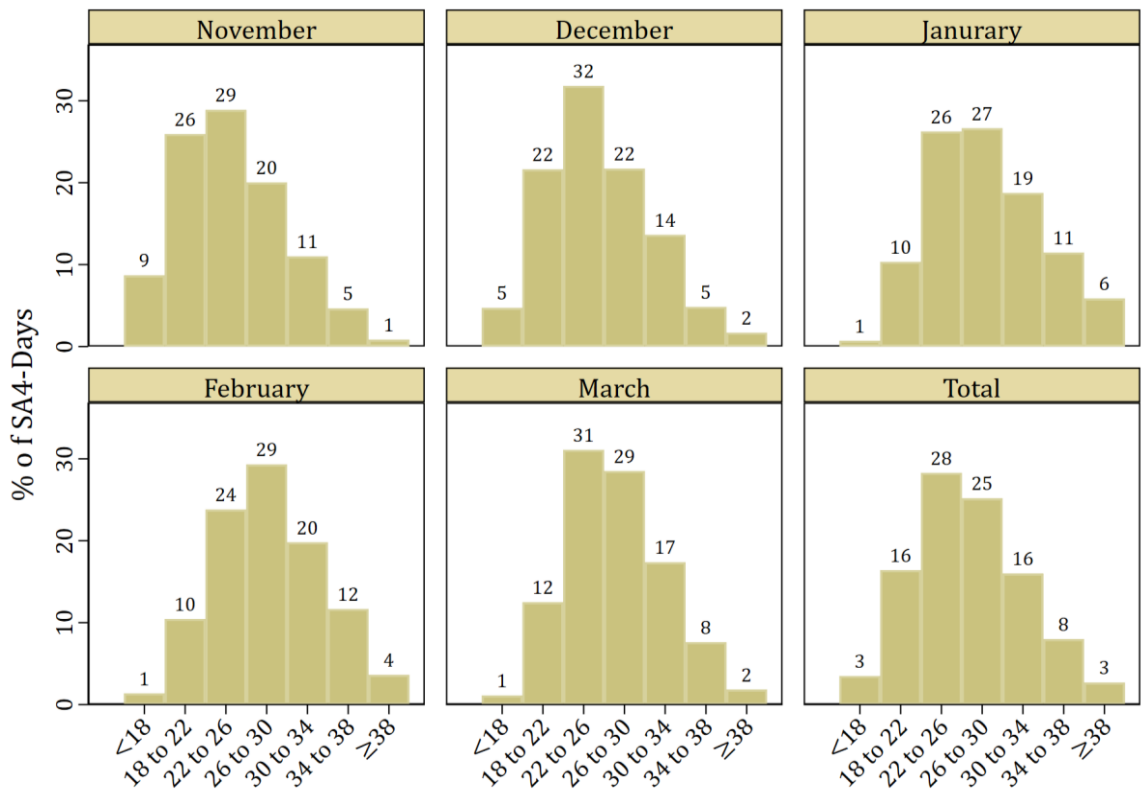

**Notes:** Based on November–March from November 2001-February 2020 for included SA4 regions.

**Figure C5. Distribution of within-individual-area interquartile range of maximum temperature**

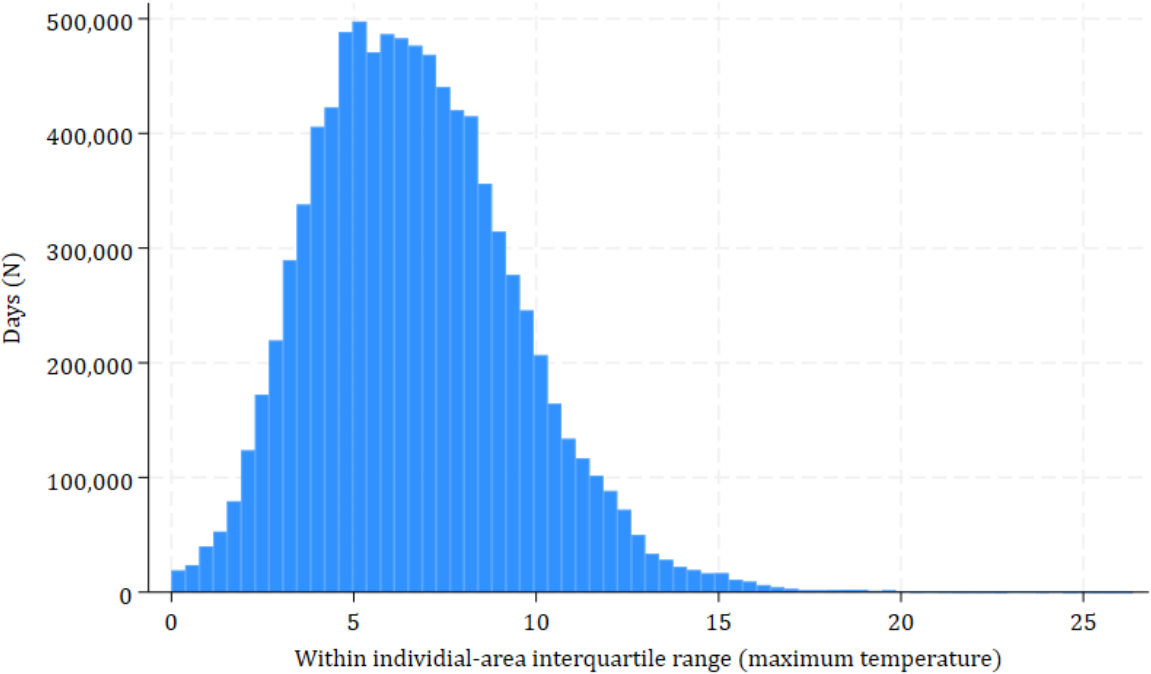

**Notes:** Interquartile range of maximum temperature calculated for each individual-area using the same sample as main results. Based on single-job holders November–March from November 2001-February 2020 for included SA4 regions. Day-of-week restricted to days the individual ever reported to attend work (i.e. Sundays only included for individuals who ever worked on a Sunday). For the period November–March from November 2001-February 2020.

# Appendix D. Robustness of Results
## Table D1. Effect of Heat on Work Attendance

|  | (1) All Workers | (2) Outdoor Workers | (3) Indoor Workers |
|---|---|---|---|
| *Panel A. Maximum Temperature* | | | |
| <22 | 0.00 | -0.38** | 0.04 |
|  | (0.07) | (0.15) | (0.07) |
| 22-26 | 0.00 | -0.33** | 0.03 |
|  | (0.05) | (0.13) | (0.05) |
| 26-30 | - | - | - |
| 30-34 | 0.01 | -0.22** | 0.04 |
|  | (0.07) | (0.11) | (0.07) |
| 34-38 | -0.05 | -0.25* | -0.04 |
|  | (0.07) | (0.14) | (0.06) |
| ≥38 | -0.96*** | -1.87*** | -0.82** |
|  | (0.12) | (0.25) | (0.13) |
| *Panel B. Maximum Wet-bulb Temperature* | | | |
| >14 | -0.10 | -0.37 | -0.07 |
|  | (0.12) | (0.26) | (0.12) |
| 14–16 | -0.00 | -0.02 | -0.00 |
|  | (0.09) | (0.17) | (0.09) |
| 16–18 | -0.05 | -0.20 | -0.03 |
|  | (0.10) | (0.17) | (0.10) |
| 18–20 | -0.04 | -0.15 | -0.03 |
|  | (0.05) | (0.13) | (0.05) |
| 20–22 | - | - | - |
| 22–24 | -0.21*** | -0.40** | -0.17** |
|  | (0.07) | (0.19) | (0.07) |
| ≥24 | -0.69*** | -1.14*** | -0.62*** |
|  | (0.21) | (0.35) | (0.22) |
| Outcome Mean | 81.17 | 81.36 | 81.13 |
| N | 9,129,807 | 856,050 | 7,998,058 |

**Notes:** Based on single-job holders November–March from November 2001-February 2020 for included SA4 regions. Day-of-week restricted to days the individual ever reported to attend work (i.e. Sundays only included for individuals who ever worked on a Sunday). Linear regression with binary outcome for daily work attendance (zero to indicate absence, 100 for attendance). Regression specification includes bins for daily maximum temperature, individual-region, month-year and day-of-year fixed effects and controls for day-of-week and precipitation. Outdoor and Indoor classification based upon occupation codes. For the period November–March from November 2001-February 2020. Sample size and outcome mean excludes singletons. Standard errors clustered by area. ***p<0.01 **p<0.05 *p<0.1

**Table D2. Effect of Heat on Work Attendance (Alternative Specifications)**

|  | (1) Main Spec. | (2) SA4 FEs | (3) Pollution | (4) Dry days only | (5) Individual Clustered Std. Err | (6) Household Clustered Std. Err |
|---|---|---|---|---|---|---|
| <22 | 0.00 | -0.03 | -0.03 | -0.05 | 0.00 | 0.00 |
|  | (0.07) | (0.07) | (0.06) | (0.09) | (0.04) | (0.04) |
| 22-26 | 0.00 | 0.02 | -0.01 | -0.01 | 0.00 | 0.00 |
|  | (0.05) | (0.04) | (0.05) | (0.07) | (0.03) | (0.03) |
| 26-30 | - | - | - | - | - | - |
| 30-34 | 0.01 | -0.02 | 0.02 | -0.00 | 0.01 | 0.01 |
|  | (0.07) | (0.06) | (0.07) | (0.07) | (0.04) | (0.04) |
| 34-38 | -0.05 | -0.08 | -0.03 | -0.01 | -0.05 | -0.05 |
|  | (0.07) | (0.06) | (0.06) | (0.07) | (0.05) | (0.05) |
| ≥38 | -0.96*** | -0.96*** | -0.92*** | -0.90*** | -0.96*** | -0.96*** |
|  | (0.12) | (0.13) | (0.12) | (0.13) | (0.09) | (0.09) |
| Individual#SA4 Fixed Effects | ✓ | ✗ | ✓ | ✓ | ✓ | ✓ |
| SA4 Fixed Effects | ✗ | ✓ | ✗ | ✗ | ✗ | ✗ |
| Air Pollution (PM2.5) | ✗ | ✗ | ✓ | ✗ | ✗ | ✗ |
| Outcome Mean | 81.17 | 81.18 | 81.17 | 81.14 | 81.17 | 81.17 |
| N | 9,129,807 | 9,140,637 | 9,129,807 | 7,169,558 | 9,129,807 | 9,129,807 |

**Notes:** Based on single-job holders November–March from November 2001-February 2020 for included SA4 regions. Day-of-week restricted to days the individual ever reported to attend work (i.e. Sundays only included for individuals who ever worked on a Sunday). Linear regression with binary outcome for daily work attendance (zero to indicate absence, 100 for attendance). Regression specification includes bins for daily maximum temperature, individual-region, month-year and day-of-year fixed effects and controls for day-of-week, precipitation and maximum wind gust. Temperatures compared to reference category of 26–30°C. Column (4) for dry days only excludes days with more than 1mm of precipitation. For the period November–March from November 2001-February 2020. Sample size and outcome mean excludes singletons. Standard errors clustered by area except where stated otherwise. ***p<0.01 **p<0.05 *p<0.1

**Figure D1. Effect of Average Daily Maximum Temperature (Usual Work Days) on Hours Worked**

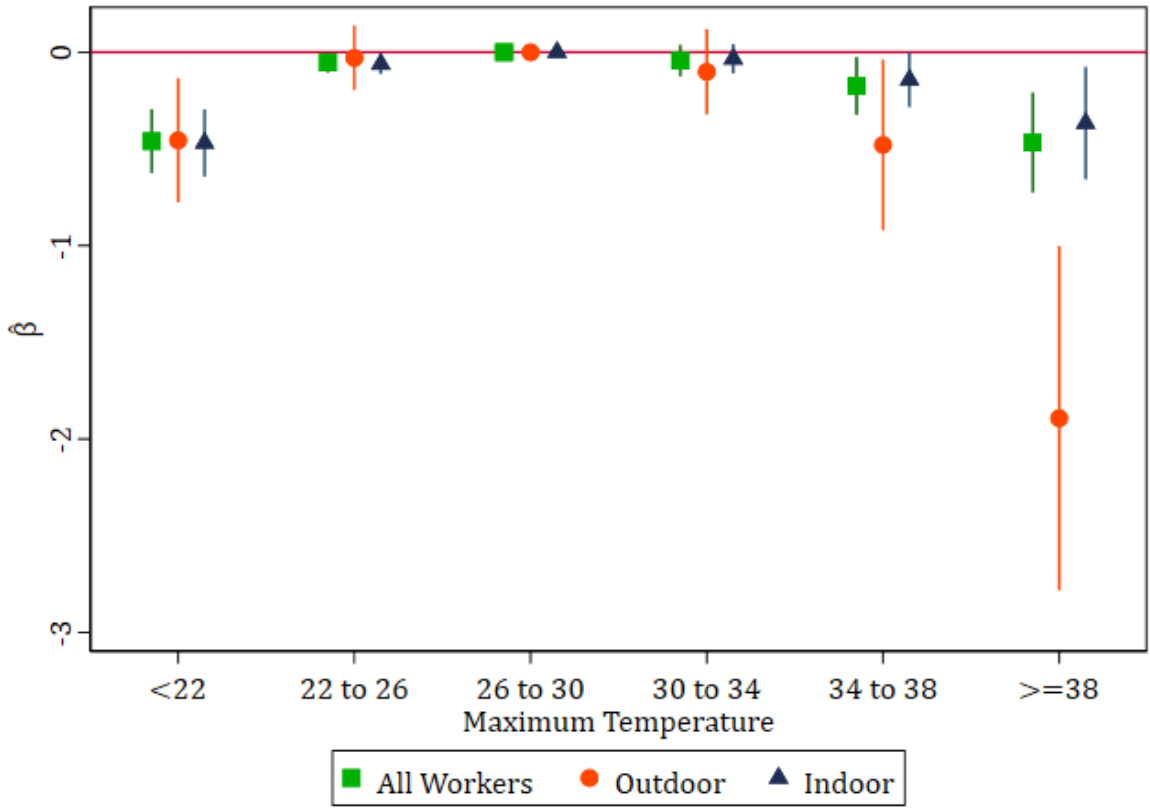

**Notes:** Based on linear regression with weekly hours worked as outcome. Regression specification includes bins for average daily maximum temperature, individual-region, month-year and week-of-year fixed effects and controls for precipitation and public holidays. Average maximum temperature denotes the average of daily maximum temperatures across usual work days. Usual work days are determined by considering the days of the week worked across all survey months. Sample includes single job holders for the period November–March from November 2001–February 2020. 95% confidence intervals, standard errors clustered by area. Outdoor and Indoor classification based on occupation codes.

**Figure D2. Effect of Temperature on Working Less than Usual Hours by Reason, Indoor Workers**

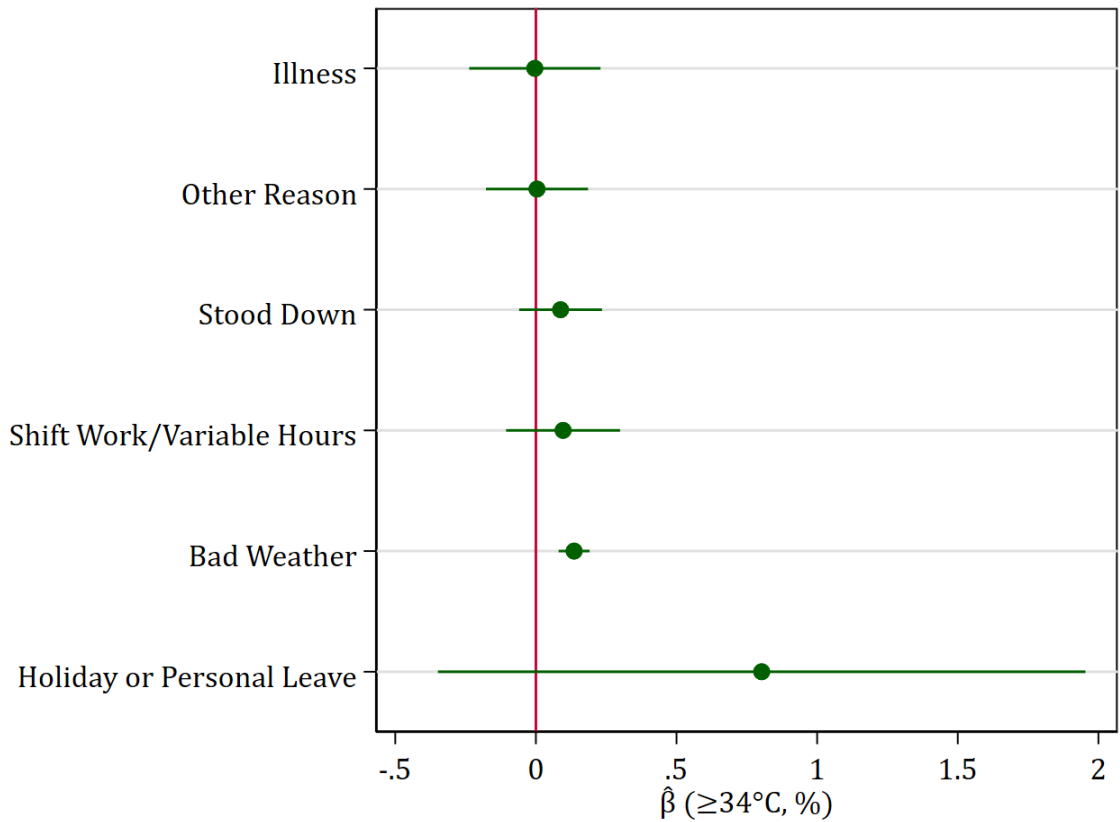

**Notes:** Based on linear regression with binary outcome indicating working less than usual hours during the week for each reason. Indoor workers only, based on occupation. Regression specification includes average daily maximum temperature ≥34°C as an indicator variable, individual-area, month-year and week-of-year fixed effects and controls for precipitation and public holidays. 95% confidence intervals, standard errors clustered by area. Outcome mean (μ) displayed on the right of chart. Sample includes single job holders for the period November–March from November 2001–February 2020.

**Figure D3. Extreme Heat and Work Attendance, Alternative Sample Restrictions**

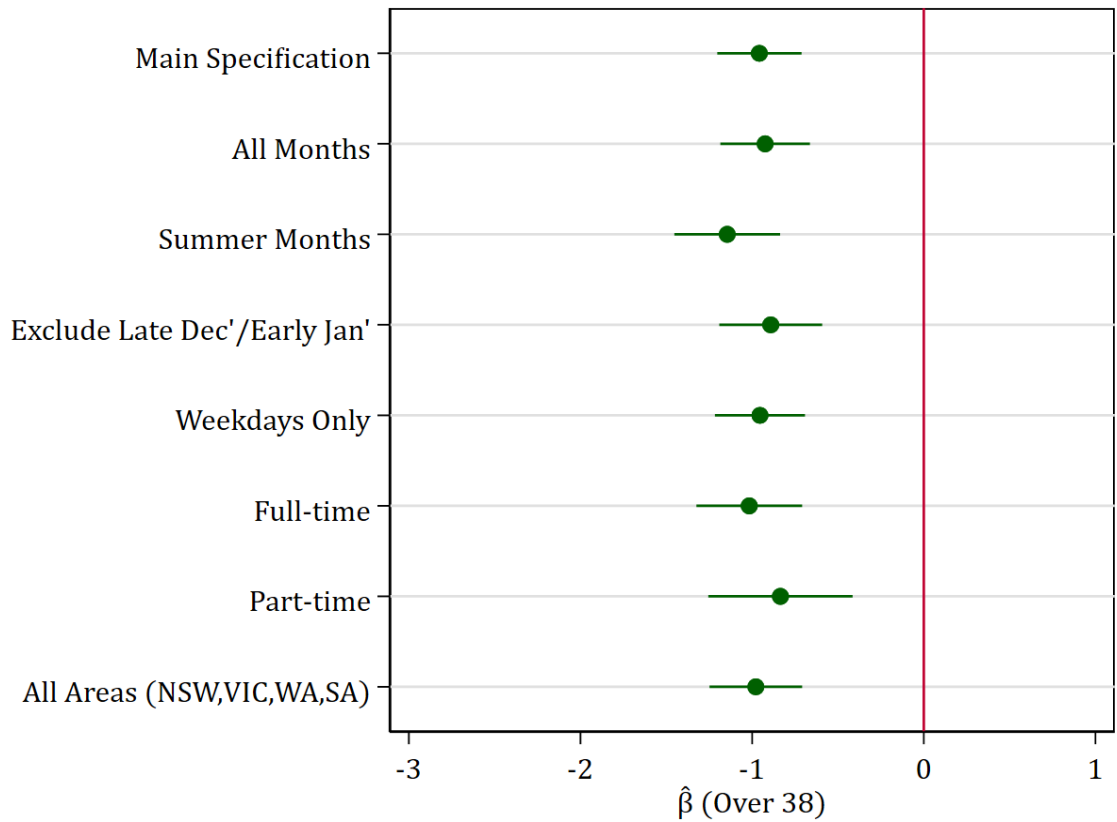

**Notes:** Based on linear regression with binary outcome for daily work attendance (zero to indicate absence, 100 for attendance). Regression specification includes bins for daily maximum temperature, individual-area, month-year and day-of-year fixed effects and controls for day-of-week and precipitation. Over 38°C compared to reference category of 26–30°C. 95% confidence intervals, standard errors clustered by area. Sample includes single job holders for the period November–March from November 2001–February 2020. Late December and early January excludes dates from the 24th December to 7th January.

**Appendix E. Consecutive Extreme Heat days**

**Table E1. Effect of Heat on Work Attendance, Consecutive Heat Days**

|  | (1) All | (2) Outdoor | (3) Indoor |
|---|---|---|---|
| Today ≥ 38°C | -1.04*** | -1.81*** | -0.92*** |
|  | (0.15) | (0.27) | (0.15) |
| Yesterday ≥ 38°C | -0.44** | -0.82*** | -0.41** |
|  | (0.18) | (0.29) | (0.19) |
| Today & Yesterday ≥ 38°C | 0.62* | 1.00 | 0.58* |
|  | (0.31) | (0.81) | (0.31) |
| Outcome mean | 81.17 | 81.35 | 81.13 |
| N | 9,237,274 | 865,298 | 8,092,027 |

**Notes:** Based on linear regression with binary outcome for daily work attendance (zero to indicate absence, 100 for attendance). Regression specification includes bins for daily maximum temperature, individual-area, month-year and day-of-year fixed effects and controls for day-of-week and precipitation. Indicator variables for maximum temperature greater than 38°C. Outdoor and Indoor classification based upon occupation codes. For the period November–March from November 2001–February 2020. Sample size and outcome mean excludes singletons. Standard errors clustered by area. ***p<0.01 **p<0.05 *p<0.1

# Appendix F. Effect of Temperature by Worker Characteristics

**Figure F1. Extreme Heat and Work Attendance by Worker Subgroup**

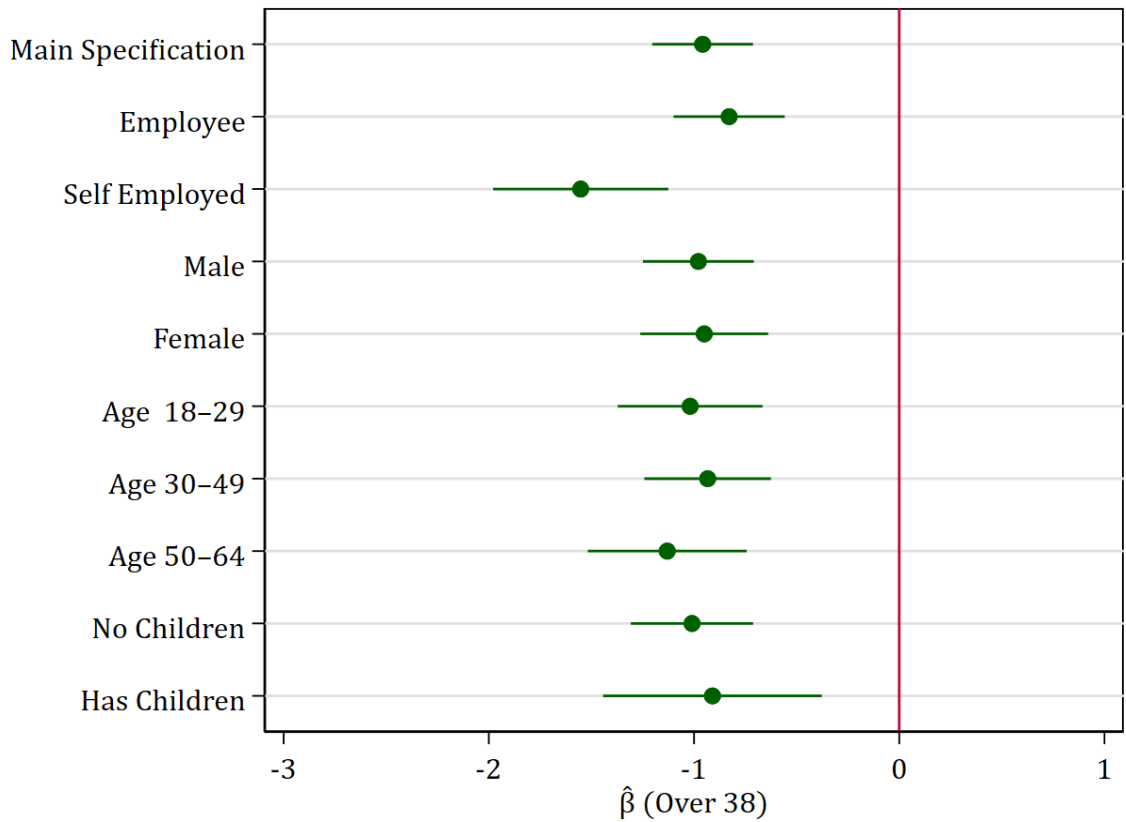

**Notes:** Based on linear regression with binary outcome for daily work attendance (zero to indicate absence, 100 for attendance). Regression specification includes bins for daily maximum temperature, individual-area, month-year and day-of-year fixed effects and controls for day-of-week and precipitation. Over 38°C compared to reference category of 26–30°C. 95% confidence intervals, standard errors clustered by area. Sample includes single job holders for the period November–March from November 2001–February 2020. Children refers to children under 15 living in the household.

**Figure F2. Effect of Extreme Temperature on Work Attendance by Occupation**

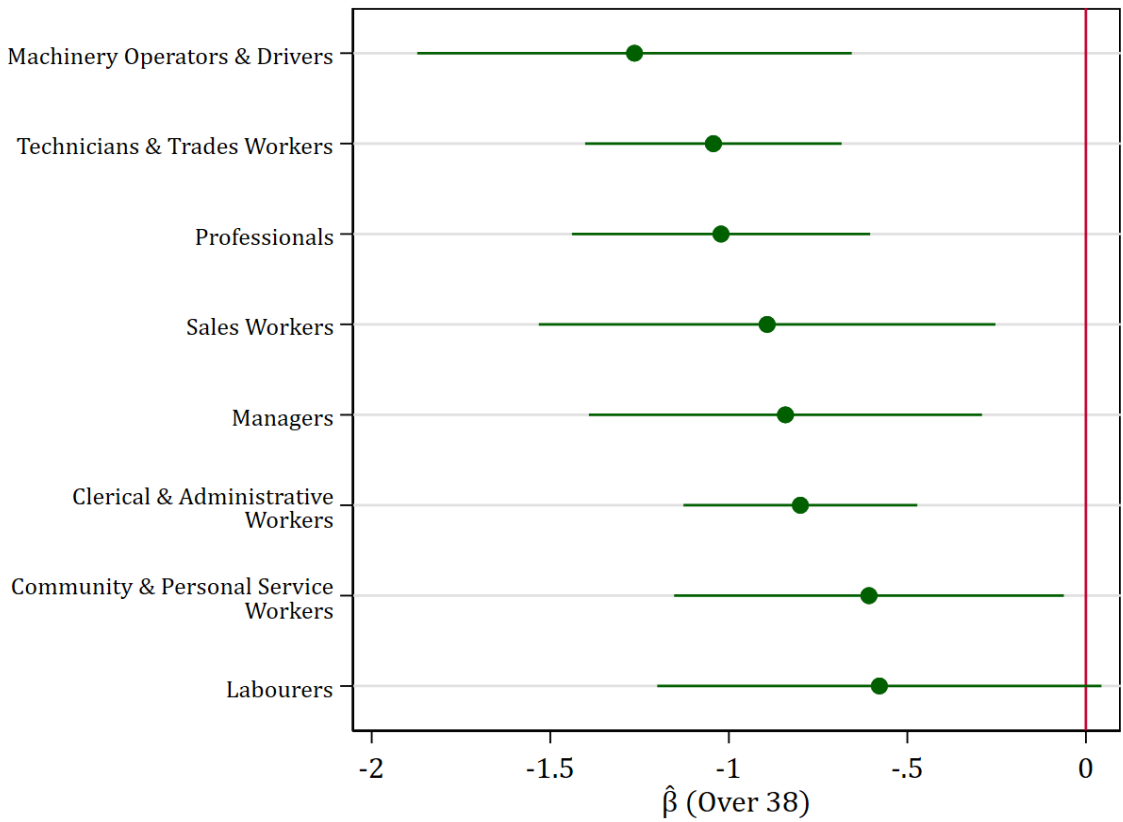

**Notes:** Based on linear regression with binary outcome for daily work attendance (zero to indicate absence, 100 for attendance). Estimated separately for each Occupation group. Regression specification includes bins for daily maximum temperature, individual-area, month-year and day-of-year fixed effects and controls for day-of-week and precipitation. Over 38°C compared to reference category of 26–30°C. 95% confidence intervals, standard errors clustered by area. For the period November–March from November 2001–February 2020.

**Appendix G. Characteristics of Industry-Occupation Groups**

We derive a "outdoor-based" score for industry-occupations using the O*NET work context data. Specifically, we use the "Outdoor, exposed to weather" characteristics to determine outdoor-based occupations (as used throughout the paper). Then, each industry-occupation group is given a score as the proportion of workers in outdoor occupations.

All other characteristics are derived using responses to the Household, Income and Labour Dynamics in Australia (HILDA) survey as follows:

- Daytime hours: Proporotion of workers with "A regular daytime schedule" as their current work schedule;
- Hours: "Hours per week usually worked in all jobs";
- Commute: Time spent travelling to and from paid employment per day worked (weekly commute time divided by average days worked per week);
- Wage: Average current weekly gross wages & salary, all jobs;
- Casual: Proportion of workers with no paid holiday leave and no paid sick leave;
- Work from home: Proportion of workers able to use "home-based work" if needed;
- Flexible hours: Proportion of "Strongly agree" and "Agree" in response to "I have a lot of freedom to decide when I do my work";
- Decide breaks: Proportion of "Strongly agree" and "Agree" in response to "I can decide when to take a break";
- Stressful job: Proportion of "Strongly agree" and "Agree" in response to "I fear that the amount of stress in my job will make me physically ill"

All measures are converted to a standard score by subtracting the sample mean and dividing by the standard deviation.

**Table G1. Effect of Temperature Exceeding 38°C on Work Attendance by Industry-Occupation**

| Industry-Occupation | $\hat{\beta}$ (Over 38°C) | Standard Error | N | Industry-Occupation | $\hat{\beta}$ (Over 38°C) | Standard Error | N |
|---|---|---|---|---|---|---|---|
| A1 | -0.06 | 1.52 | 56,900 | J5 | -1.36 | 1.23 | 36,060 |
| A8 | -2.33 | 1.16 | 32,768 | K1 | -1.49 | 0.93 | 56,217 |
| B2 | -2.48 | 1.25 | 31,779 | K2 | -2.53 | 0.53 | 132,713 |
| B3 | -0.19 | 1.60 | 30,433 | K5 | -1.62 | 0.62 | 151,272 |
| B7 | -0.79 | 2.44 | 32,934 | L5 | 1.45 | 1.12 | 33,740 |
| C1 | -0.67 | 0.51 | 125,750 | L6 | 0.29 | 0.97 | 68,075 |
| C2 | 0.21 | 0.85 | 85,985 | M1 | -1.20 | 0.60 | 88,306 |
| C3 | -1.37 | 0.38 | 241,974 | M2 | -1.09 | 0.37 | 435,776 |
| C5 | -0.20 | 0.67 | 89,705 | M3 | -0.67 | 0.59 | 75,144 |
| C6 | 0.56 | 1.27 | 37,308 | M5 | -1.81 | 0.74 | 130,486 |
| C7 | -1.31 | 0.56 | 120,148 | N1 | -0.62 | 1.02 | 35,193 |
| C8 | -1.01 | 0.65 | 152,424 | N2 | -0.51 | 0.94 | 42,619 |
| E1 | 0.06 | 1.00 | 77,204 | N4 | 0.79 | 1.49 | 38,686 |
| E3 | -1.40 | 0.25 | 427,739 | N5 | -0.52 | 1.16 | 57,072 |
| E5 | -1.66 | 0.84 | 71,628 | N8 | -2.09 | 0.80 | 112,811 |
| E7 | -2.81 | 1.13 | 52,771 | O1 | 0.29 | 0.95 | 81,338 |
| E8 | -2.40 | 0.71 | 121,830 | O2 | -1.39 | 0.69 | 169,067 |
| F1 | -1.83 | 0.65 | 81,254 | O3 | -1.09 | 1.53 | 31,260 |
| F2 | -0.58 | 0.89 | 43,412 | O4 | -1.00 | 0.88 | 115,345 |
| F5 | -0.71 | 0.86 | 71,376 | O5 | -0.52 | 0.46 | 186,397 |
| F6 | 0.58 | 0.74 | 58,987 | P1 | -1.83 | 0.99 | 44,152 |
| F7 | -2.69 | 0.97 | 53,439 | P2 | -0.49 | 0.42 | 361,088 |
| G1 | -1.57 | 0.62 | 168,027 | P4 | -0.46 | 1.24 | 60,733 |
| G2 | -0.40 | 1.14 | 40,223 | P5 | 0.26 | 1.02 | 64,554 |
| G3 | -1.99 | 0.79 | 56,022 | Q1 | -1.64 | 1.10 | 47,127 |
| G5 | -2.41 | 1.05 | 66,919 | Q2 | -0.90 | 0.45 | 392,805 |
| G6 | -1.12 | 0.56 | 491,649 | Q3 | -1.52 | 1.00 | 35,721 |
| G7 | -0.67 | 1.34 | 47,052 | Q4 | -0.49 | 0.53 | 286,318 |
| G8 | 1.91 | 1.16 | 65,551 | Q5 | -0.09 | 0.68 | 136,561 |
| H1 | -1.02 | 0.96 | 114,178 | Q8 | 1.67 | 1.15 | 50,937 |
| H3 | 0.22 | 0.81 | 91,138 | R2 | 1.64 | 1.72 | 35,902 |
| H4 | 0.25 | 0.83 | 152,849 | R4 | -1.55 | 1.18 | 48,114 |
| H6 | 1.23 | 1.04 | 91,588 | S1 | 0.64 | 1.19 | 30,141 |
| H8 | 0.25 | 0.82 | 120,675 | S2 | -1.95 | 1.26 | 36,460 |
| I1 | -0.57 | 1.26 | 42,669 | S3 | -1.12 | 0.70 | 173,285 |
| I5 | -0.72 | 0.61 | 101,278 | S4 | -2.29 | 1.30 | 37,863 |
| I7 | -1.37 | 0.44 | 214,838 | S5 | 0.32 | 0.99 | 41,492 |
| J2 | -1.71 | 1.07 | 72,709 | S8 | -0.69 | 1.30 | 37,589 |
| J3 | -0.21 | 1.13 | 34,342 | | | | |

**Notes:** Industry based on ANZSIC 2006 divisions (A: Agriculture & Forestry; B: Mining; C: Manufacturing; D: Energy, Water & Waste; E: Construction; F: Wholesale; G: Retail; H: Hospitality; I: Transport & Warehousing; J: Media & Communication; K: Finance & Insurance; L: Real Estate; M: Professional &Scientific; N: Admin & Support; O: Public Admin & Safety; P: Education; Q: Health; R: Arts & Recreation; S: Other Services). Occupation based on ANZSCO 2006 major groups (1: Managers; 2: Professionals; 3: Technicians & Trades; 4: Community & Personal Service; 5: Clerical & Administrative; 6: Sales; 7: Machinery Operators and Drivers; 8: Labourers).

**Table G2. Importance of work characteristics on the effect of extreme heat on work attendance**

| | | |
|---|---:|---:|
| Outdoor | -0.06 | (0.14) |
| Daytime schedule | -0.07 | (0.14) |
| Commute time | -0.24* | (0.13) |
| Casual (no paid leave) | -0.20 | (0.22) |
| Casual (no paid leave) × outdoor | -0.10 | (0.14) |
| Casual (no paid leave) × commute time | -0.26** | (0.13) |
| Earnings | -0.23 | (0.18) |
| Work from home | 0.13 | (0.21) |
| Flexible hours | 0.33 | (0.21) |
| Decide breaks | -0.36 | (0.27) |
| Job stress | 0.12 | (0.14) |

**Notes:** Based on largest industry-occupation groups (N=77). Linear regression with the estimated heat-effect as the dependent variable (over 38°C on work attendance), weighted by group sample size (from Labour Force Survey). Outcome mean is -0.78. Job characteristics for industry-occupation groups are based on O*NET and responses to the HILDA survey (see Appendix G for details). All regressors are standardized by demeaning and dividing by the standard deviation for comparison. Standard errors in parentheses. ***p<0.01 **p<0.05 *p<0.1

**Appendix H. Public Transport Usage and Performance**

We use publicly available data on public transport usage and performance from New South Wales (Transport for NSW, 2023b). We investigate the effect of temperature on three outcome variables: (i) monthly train station entry and exits; (ii) monthly bus trips and; (iii) daily train punctuality as a percentage of on trips. We use the location of the train station, the centroid of the bus contract zone and the middle station along the train line to match with temperature for the three outcomes respectively.

For outcomes (i) and (ii), which are monthly, we use the following specification:

$$\text{Ln}(Trips_{it}) = \beta_1 DaysBelow22_{it} + \beta_2 Days22to26_{it} + \beta_3 Days30to34_{it} + \beta_4 Days34to38_{it} + \beta_5 DaysAbove38_{it} + \tau_{it}$$

The temperature coefficients, $\beta_1 - \beta_5$, can be interpreted as the impact of an additional day within the month, relative to 26–30°C. We include location-year and location-month fixed-effects, $\tau_{it}$.

Train punctuality, outcome (iii), is recorded daily as the percentage of trains on time. We use the following specification:

$$OnTime_{it} = \sum_{j=1}^{J} \boldsymbol{\beta} * temp_{j,it} + \tau_{it}$$

We use the same 4°C-bins used throughout the paper with a reference category of 26–30°C. We use location-year-month fixed effects, $\tau_{it}$.